\documentclass[aps,prd,twocolumn,a4paper,superscriptaddress,10pt]{revtex4-1}
\usepackage{graphicx}
\usepackage{multirow}
\usepackage{latexsym}
\usepackage{hyperref}
\usepackage[british]{babel}
\usepackage{amsmath}
\usepackage{wrapfig}
\usepackage{amsmath}
\usepackage{xcolor}

\begin{document}

\title{High resolution calibration of the cosmic strings velocity dependent one-scale model}
\author{J. R. C. C. C. Correia}
\email{Jose.Correia@astro.up.pt}
\affiliation{Centro de Astrof\'{\i}sica da Universidade do Porto, Rua das Estrelas, 4150-762 Porto, Portugal}
\affiliation{Instituto de Astrof\'{\i}sica e Ci\^encias do Espa\c co, Universidade do Porto, Rua das Estrelas, 4150-762 Porto, Portugal}
\affiliation{Faculdade de Ci\^encias, Universidade do Porto, Rua do Campo Alegre 687, 4169-007 Porto, Portugal}
\author{C. J. A. P. Martins}
\email{Carlos.Martins@astro.up.pt}
\affiliation{Centro de Astrof\'{\i}sica da Universidade do Porto, Rua das Estrelas, 4150-762 Porto, Portugal}
\affiliation{Instituto de Astrof\'{\i}sica e Ci\^encias do Espa\c co, Universidade do Porto, Rua das Estrelas, 4150-762 Porto, Portugal}

\date{24 April 2021}

\begin{abstract}
The canonical velocity-dependent one-scale (VOS) model for cosmic string evolution must be calibrated using high resolution numerical simulations, We exploit our state of the art graphics processing unit accelerated implementation of the evolution of local Abelian-Higgs string networks to provide a detailed and statistically robust calibration of the VOS model. We rely on the largest set of high resolution simulations carried out to date, with a wide range of cosmological expansion rates, and explore the impact of key numerical parameters, including the dynamic range (comparing box sizes from $1024^3$ to $4096^3$), the lattice spacing, and the choice of numerical estimators for the string velocity. We explore the sensitivity of the VOS model parameters to these numerical parameters, with a particular emphasis on the observationally crucial loop chopping efficiency, and also identify key differences between the equation of state and conjugate momentum estimators for the string velocities, showing that the latter one is more reliable for fast expansion rates (while in Minkowski space the opposite has been previously shown). Finally, we briefly illustrate how our results impact observational constraints on cosmic strings. 
\end{abstract}
\maketitle
\allowdisplaybreaks

\section{\label{intr}Introduction}

Topological defects are possible fossil relics of the early Universe, encoding information on the physical conditions therein. They form as a consequence of symmetry breaking phase transitions, via the Kibble mechanism \cite{Kibble:1976sj}, and if stable they will persist until the present day. Depending on the details of the symmetry broken (specifically, on the homotopy group of the vacuum manifold), different types of defects can form, with different dimensionalities. One type of defect which is generic enough to form in many candidate theories of physics beyond the standard model \cite{Jeannerot:2003qv,Sarangi:2002yt} are the benign one-dimensional line-like cosmic strings---benign in the sense that they cannot overclose the Universe, at least in the simplest models. Given how generic these fossil relics are, they are a primary target for constraints from current observational facilities \cite{LIGODefects,PlanckDefects}. For these observational studies, detection of strings would signal new theories of physics beyond the Standard Model, while a non-detection would enable constraints on the mass-scale of strings, which is directly related to the symmetry breaking scale. However, many approximations are done in the current analyses, meaning that the current constraints are not fully reliable, and sometimes small changes in the assumptions underlying the analysis lead to derived constraints differ by several orders of magnitude. In other words, the systematic uncertainties are far larger than the statistical ones.

The evolution of cosmic string networks is highly non-linear, and must be studied by a combination of analytic modelling and numerical simulations \cite{Book}.That said, there is at least one clear feature of the simplest models of cosmic string networks that is revealed by both analytical and numerical studies: in cosmological epochs where the scale factor is proportional to some power of the physical time, the attractor (i.e., asymptotic) behaviour is known as the scaling (or scale invariant) regime, where the network's rate of separation of strings and the velocity are asymptotically constant. This assumption is computationally useful: even if simulations cannot cover all of the observationally relevant cosmological evolution, their behaviour can be extrapolated to all cosmological epochs and transitions therein by means of a simulation with a limited dynamic range together with a properly calibrated model.

The canonical analytic evolution model for cosmic strings and other topological defects is the Velocity dependent One-Scale (henceforth VOS) model, originally developed for cosmic strings \cite{Martins:1996jp,Martins:2000cs} and subsequently extended to other defects---see \cite{Book} for a recent review. On the other hand, there are two main types of simulations of strings---Nambu-Goto \cite{AS,BB,Blanco} and field theory \cite{Bevis:2006mj,Hindmarsh:2017qff}---which confirm the presence of the scale invariant solution but significantly disagree on the details. This is relevant because the observational consequences of string networks directly depend on these details, including on the means by which this scaling solution is dynamically reached and sustained, and more specifically on the energy loss mechanisms responsible for doing so.

Recently we have started a systematic program aiming to obtain a full and statistically robust calibration of an extension of the VOS model. This extended model incorporates new parameters to explicitly account for the correct velocity dependencies of both a generalized curvature term and an explicit term for modelling radiative losses, including contributions from loop production and from scalar and gauge radiation. This relies on a new generation graphics processing unit (GPU) accelerated evolution code for Abelian-Higgs cosmic strings \cite{Correia:2018gew,Correia:2020yqg} which has been recently shown to be more than 30 times faster than the best previously available code and enables the collection of statistically significant data sets using manageable amounts of computing resources and wall clock time.

In \cite{Correia:2019bdl}, henceforth Paper 1, we have presented the extended version of the VOS model and provided a preliminary calibration thereof, relying on more than one thousand $512^3$ simulations for a wide range of different cosmological expansion rates. In \cite{Correia:2020gkj}, henceforth Paper 2, we studied the sensitivity of the model calibration to the presence (or absence) of thermal oscillations due to high gradients in the initial conditions (showing that a small amount of cooling has no statistically significant impact on the VOS model calibration, while a longer dissipation period does have a noticeable effect) and also introduced an improved Markov Chain Monte Carlo (MCMC) based pipeline for calibrating the VOS model. The present work is the continuation of this program. We rely on the largest set of high resolution simulations gathered to date, with box sizes from $1024^3$ to $4096^3$ and with a wide range of cosmological expansion rates, to provide a more precise calibration of the VOS model. In doing so, we also explore the impact of key numerical parameters: in addition to the dynamic range (which is primarily related to the box size), we also investigate the effects of lattice spacing and the choice of numerical estimators of the string mean velocity.
 
This rest of the work is organized as follows. We start in Sect. \ref{pre} by briefly describing the VOS model, its parameters and main features, as well as the field theory simulations that we use and the simulation diagnostics that are used to calibrate the VOS model. We then proceed to describe our detailed calibration and its sensitivity to the relevant numerical parameters in Sects. \ref{cal} and \ref{lat}, finding that the VOS model parameter that is more sensitive to these choices is the loop chopping efficiency, and also characterizing key differences between two alternative estimators for the string velocities. Finally, in Sect. \ref{cls} we briefly illustrate how our results impact observational constraints on cosmic strings, and present some conclusions in Sect. \ref{conc}. 

\section{\label{pre}Prelude}

We start by presenting brief overviews of the extended VOS model and of our Abelian-Higgs field theory numerical simulation code, including our estimators for the string characteristic length scale and velocity. Our goal here is not to be exhaustive but rather to present a concise introduction of the concepts that will be relevant for the results discussed in the rest of the work. We refer the reader to the cited references for more detailed discussions.

\subsection{\label{VOS}Extended Semi-analytical modelling}

The Velocity-dependent One-Scale (VOS) model \cite{Martins:1996jp,Martins:2000cs} comprises two coupled differential equations, describing how two network averaged quantities---the mean correlation length or characteristic length scale and the root mean squared velocity---evolve over cosmic time. It can thus be thought of as a thermodynamic model. In physical coordinates the model can written as,
\begin{equation}
  2\frac{dL}{dt} = 2HL(1+v^2) + F(v)
\end{equation}
\begin{equation}
  \frac{dv}{dt} = \bigg( 1-v^2 \bigg) \bigg( \frac{k(v)}{L} -2Hv \bigg)
\end{equation}
where $L$ is the mean string correlation length, $v$ the velocity, and $H$ the Hubble parameter.

There are also two functions which depend explicitly on the velocity: the momentum parameter $k(v)$, and the energy loss function $F(v)$. In the standard VOS model, the phenomenological parameter $k(v)$ has a form derived from considering the helicoidal string solution and by comparison with Nambu-Goto simulations in non-relativistic and relativistic regimes \cite{Martins:1996jp,Martins:2000cs}. Later, work based on field theory simulations of domain walls and leading to an accurate VOS walls model \cite{Rybak1} led to a more general form of this function, where certain parameters would be fixed by direct calibration from simulations. This has been adapted to the case of cosmic strings in Paper 1, and has the form
\begin{equation}
  k(v) =  k_0\frac{1-(qv^2)^{\beta}}{1+(qv^2)^{\beta}}.
\end{equation}
where of the three free parameters ($k_0$, $\beta$ and $q$) two have a clear physical meaning: $k_0$ is the maximal value of the momentum parameter (i.e., its low-velocity limit) and $1/q$ can at most be equal to the maximum string velocity squared; on the other hand $\beta$ is a more phenomenological parameter allowing for a generic power law dependence.

The second velocity-dependent function $F(v)$ is physically an energy loss function. In the original VOS model, it merely encapsulated how the string network would lose energy over time via loop production, i.e. the Kibble's loop chopping efficiency. In the extended VOS, $F(v)$ is modified to include an additional radiative term, as follows 
\begin{equation}
  F(v) = cv +  d[k_0-k]^r\,.
\end{equation}
Here we have three additional free parameters: $c$ and $d$ are the normalization factors for the contributions from loop chopping efficiency and from scalar and gauge radiation components (massive and massless), and $r$ quantifies a power law of the curvature parameter---see \cite{Rybak1} for a detailed justification of this assumption.

Apart from having been used in the domain walls case \cite{Rybak1}, a preliminary calibration on the six model parameters can be found in Paper 1 and Paper 2. An important difference between the two cases, highlighted in these previous works, is that for domain walls the radiation losses term always dominates over loop production, but this is not the case for cosmic strings. We will return to this point in what follows.

The main goal of the present work is to improve the earlier calibration of the extended VOS model, while quantifying possible sources of systematic errors and biases. For the comparison of the model to simulations it is convenient to replace the physical time $t$, physical correlation length $L$ and Hubble parameter $H$ by their co-moving counterparts, respectively $\eta$, $\xi$, and $\mathcal{H}$. This is done because the simulations are evolved in comoving coordinates. This leads to
\begin{equation}
  \label{eq:vos1_comoving}
  \frac{d\xi}{d\eta} = \frac{m\xi}{(1-m)\eta} v^2 +F(v)
  \end{equation}
  \begin{equation}
  \label{eq:vos2_comoving}
  \frac{dv}{d\eta} = (1-v^2) \bigg[ \frac{k(v)}{\xi}  - \frac{2m v}{(1-m)\eta} \bigg],.
\end{equation}

Note that since in what follows we will simulate expanding universes where the scale factor obeys the power law $a \propto t^m \propto \eta^{m/(1-m)}$, we already substituted the appropriate expression for $\mathcal{H}$ in terms of $m$.

In this form the model is ready for calibration. Underlying the calibration procedure is the assumption that the model contains a fixed point solution for any power law universe being simulated---the aforementioned scale invariant solution. Such a behavior is described by the following relations
\begin{align}
  \xi \propto \eta  \propto d_H && v = const.
\end{align}
where $d_H$ is the horizon size. The attractor nature of this solution is analytically well known \cite{Book}. The statistical comparison between simulations of string in such Universes and the VOS model will be done using the same MCMC based pipeline as described in Paper 2, which includes automatic uncertainty propagation and minimization coupled with Bayesian inference to explore the model's parameter space. 

\begin{figure}
\centering
  \includegraphics[width=1.00\columnwidth]{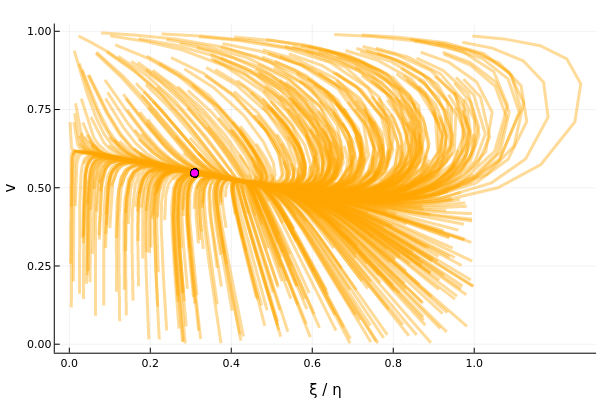}
  \caption{600 trajectories of the extended VOS model in phase space. Each trajectory is obtained by evolving some random initial condition for the velocity ($v$) and the correlation length divided by conformal time ($\xi / \eta$) with the extended model calibrated with $4096^3$ radiation era boxes, lattice spacing $\Delta x = 0.5$, equation of state velocity estimator, through the conformal time range $[1.0, 512.0]$ (see the main text for details on these parameters). We also mark the last time of the integration for every solution with a magenta dot, to indicate the position of the fixed point.}
  \label{fig1}
\end{figure}

In addition, we will also invert the VOS equations to obtain numerically measured values of the momentum parameter and the energy loss function in the scaling regime,
\begin{equation}\label{eq:vos1}
  F(v) = 2 \epsilon [1-m (1+v_0^2)]
\end{equation}
\begin{equation}\label{eq:vos2}
  k(v) = 2m\epsilon v_0
\end{equation}
where $\epsilon$ is given by $\xi/(\eta(1-m))$ at scaling, and $v_0$ is the velocity in the same regime. Note that in practice, and due to the choices of initial conditions, $\epsilon$ will instead be numerically given by $\xi/((\eta-\eta_0)(1-m))$ where $\eta_0$ is a numerical offset of no physical significance. Readers interested in this technical (numerical) point can find a discussion in Paper 2.

We note that even though the fixed point nature of all possible scaling solutions in the extended model has not yet been thoroughly studied via a rigorous dynamical systems analysis, one can show that this is indeed the case for the model parameters of interest. Specifically, this can be seen in Fig. \ref{fig1}, where a set of 600 trajectories of the VOS model, each with different initial conditions and using the calibration from $4096^3$ simulations (to be discussed in what follows) is used, shows the fixed point clearly. We leave a more detailed exploration of this phase space to a follow-up publication.

\subsection{\label{setup}Simulation setup}

There are two methods for simulating cosmic string networks. The first evolves string segments in the infinitely thin string limit, also known as the Nambu-Goto approximation \cite{BB,AS,Fractal,VVO,Blanco}. The second one evolves fields on a comoving lattice, and the field configuration arises naturally from topological considerations and Hubble damping \cite{Moore:2001px,Bevis:2006mj,Hindmarsh:2017qff}. Our GPU-accelerated simulations \cite{Correia:2018gew,Correia:2020yqg} are of the latter type. We will use throughout this work a maximum of $4096$ graphical accelerators of Piz Daint supercomputer, the 12th most powerful supercomputer at the time of writing \cite{TOP}. 

In order to describe a field theory string simulation, consider a Lagrangian density of the following form
\begin{equation}
  \mathcal{L}=|D_\mu \phi|^2 - \frac{\lambda}{4}(|\phi|^2 -1)^2 - \frac{1}{4e^2}F^{\mu \nu}F_{\mu \nu}\,,
\end{equation}
where $\phi$ and $A$ are the scalar and gauge fields of the theory, $\lambda$ and $e$ are their corresponding coupling constants, $F_\mu \nu$ is the gauge field strength and $D_\mu$ indicates gauge derivatives. An Abelian-Higgs string simulation typically starts with some random initial conditions (mimicking the fields after the symmetry breaking in a computationally cheap way) and evolves this field configuration forward, comoving timestep by timestep, stopping when the horizon reaches the box size (at which point the periodic boundary conditions in the simulation box are no longer representative of an expanding universe). In our case, the initial conditions correspond to random phases of the complex scalar field, with all other fields set to zero. Note that we could attempt to apply an initial period of cooling, however our goal in this manuscript is not to understand how cooling can affect model calibration---this has already been done in Paper 2.

The evolution of the fields is described by a discrete form of the following equations of motion
\begin{equation}
  \ddot{\phi} + 2\frac{\dot{a}}{a}\dot{\phi} = D^jD_j\phi -\frac{a^{2}\lambda}{2} (|\phi|^2 - 1) 
\end{equation}
\begin{equation}
  \dot{F}_{0j} = \partial_j F_{ij} -2a^2 e^2 Im[\phi^* D_j \phi]\,.
\end{equation}
where the couplings $\lambda$ and $e$ are related to their physical counterparts by a factor of $a^{(1-\kappa)}$, and the choice $\kappa=0$ forces the string to have a constant comoving width, also known as the Press-Ryden-Spergel approximation \cite{PRS}. On the other hand the choice $\kappa=1$ allows the strings to shrink in comoving coordinates. While the latter option is the expected physical behavior, in what follows we set $\kappa=0$ in order to avoid tuning a core growth phase (setting a negative $\kappa$) every time the simulation is run with a different expansion rate. Note that core growth---as done in \cite{Bevis:2006mj}---followed by  physical behavior would have significant disadvantages, which have been discussed in Paper 1.

For our calibration of the VOS model, we need to output two mean quantities: the mean string separation (interchangeably the mean correlation length), and the mean velocity squared. In our simulation the two naturally available correlation length estimators were shown to produce comparable results, differing at most by a few percent at very high expansion rates as discussed in Paper 1. In what follows we use the Winding length estimator, which can be shown to yield the mean string length $\xi_W$ through the
\begin{equation}
  \label{eq:defxiW}
  \xi_W = \sqrt{\frac{\mathcal{V}}{\sum_{ij,x} W_{ij,x}}}\,,
\end{equation}
where $\mathcal{V}$ is the box volume and the total length of string in the box corresponds to a summation of the non-zero windings found piercing the cell faces throughout the lattice. The gauge-invariant winding $W_{i,j}$ at lattice site $x$ is defined, as shown by \cite{Kajantie:1998bg}, as
\begin{equation}
  W_{ij} = \frac{1}{2\pi} (Y_{i,x} + Y_{j,x+i} - Y_{i,x+j} -  Y_{j,x})\,,
\end{equation}
  where $Y_i$ is given by
\begin{equation}
  Y_i = [(\phi^x)_{arg} -(\phi^{x+k_i})_{arg} +A_{i,x}  ]_\pi - A_{i,x}\,.
\end{equation}
and when $W_{ij}$ is non-zero at a particular plaquette it indicates the presence of a straight string segment of length $\Delta x$. 

Similarly, two possible velocity estimator are available in our simulation, one based on Lorentz boosting a static straight string, and another one based on the equation of state. Both are derived and described in detail in \cite{Hindmarsh:2017qff}, and as such we will merely state their definitions here. The first estimator is given by
\begin{equation}
  \label{eq:defvPhi}
  <v^2>_{\phi} = \frac{2R}{1+R}\,,
\end{equation}
  where $R$ is given by
\begin{equation}
  \label{eq:defR}
  R = \frac{\sum_x |\Pi|^2 \mathcal{W}}{\sum_{x,i} |D^+_{x,i} \phi|^2 \mathcal{W} }
\end{equation}
and $\Pi$ is the conjugate momentum of the scalar field. Here $\mathcal{W}$ is a weight function meant to localize the estimators around the strings, which in what follows will be given by the Lagrangian density. The second velocity estimator option is given by
\begin{equation}
  \label{eq:defEoS}
  <v^2>_{\omega} = \frac{1}{2} \bigg( 1+3\frac{\sum_x p_x \mathcal{W}_x}{\sum_x \rho_x  \mathcal{W}_x} \bigg)\,;
\end{equation}
where the equation of state parameter $\omega$ is computed by the box average density and pressure (both of them being appropriately weighted by the Lagrangian density). While in the case of the string length scale estimators the choice between the two available options has very little impact in the calibration, we will see that the choice of velocity estimator does heavily impact the calibrated parameters, with the difference being ascribed to the different behaviours of the two estimators at high expansion rate and, having identified the issue, we also explore how to address this systematic error.

\begin{figure*}
\centering
  \includegraphics[width=1.0\columnwidth]{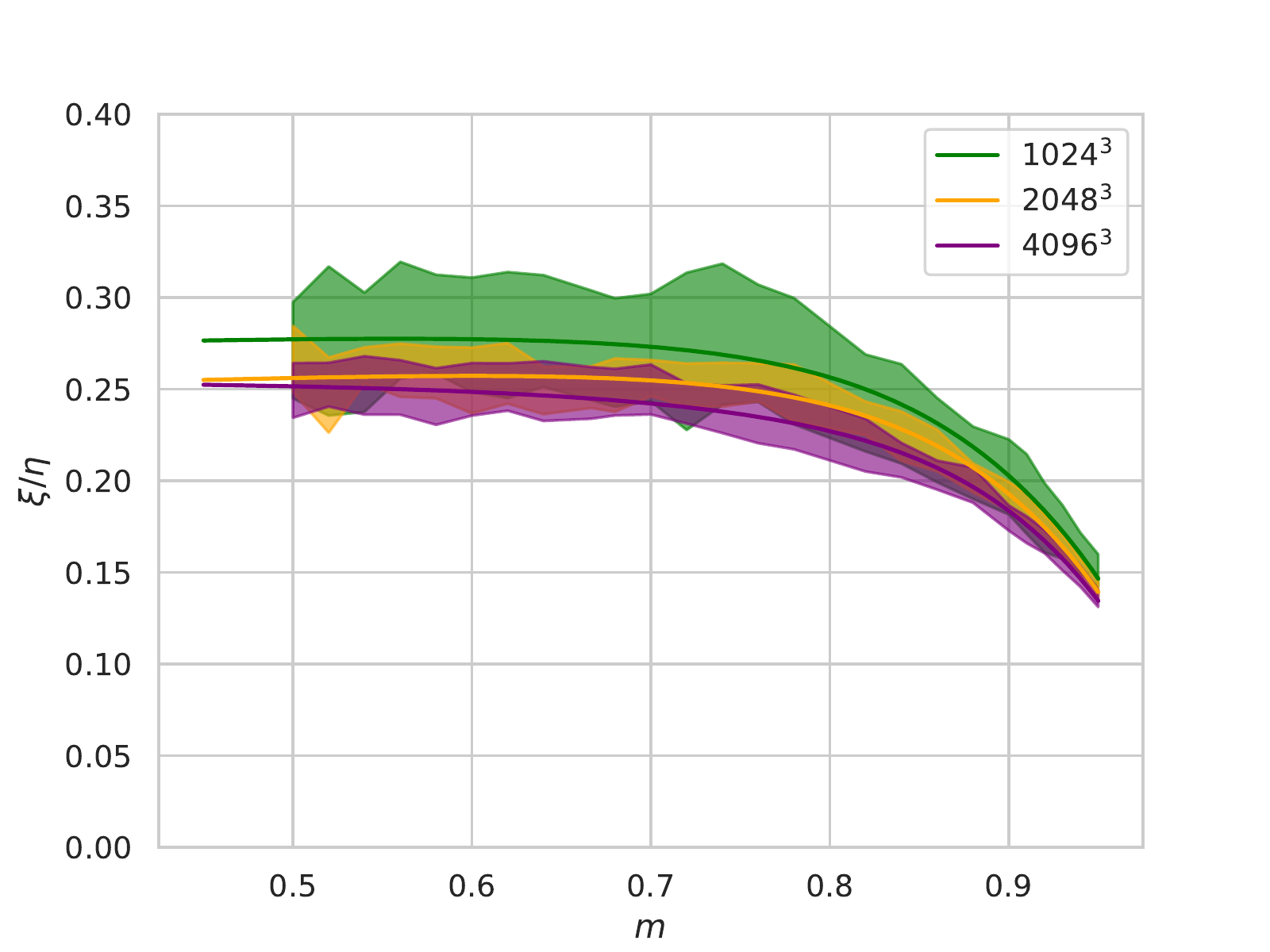}
  \includegraphics[width=1.0\columnwidth]{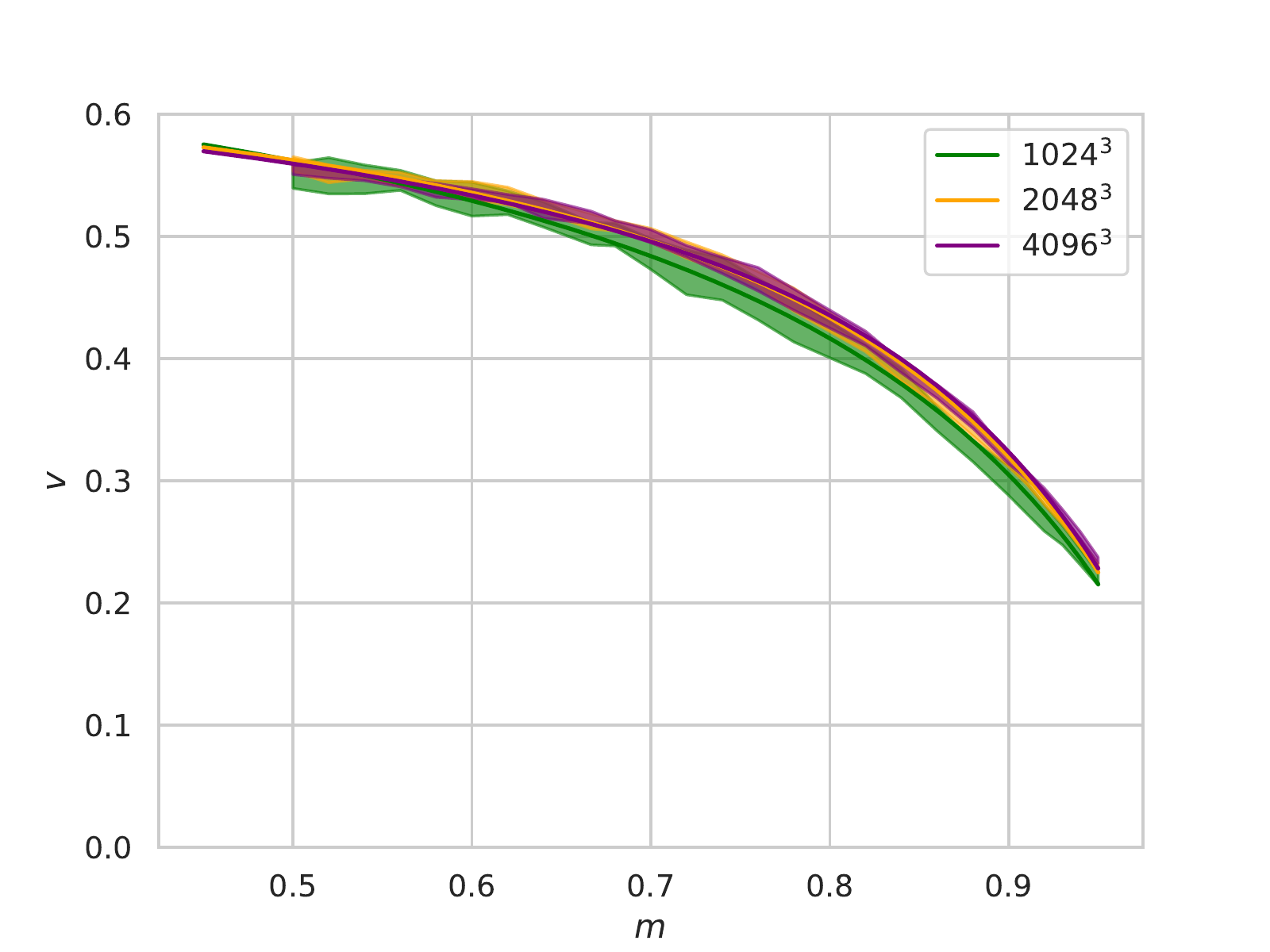}
  \includegraphics[width=1.0\columnwidth]{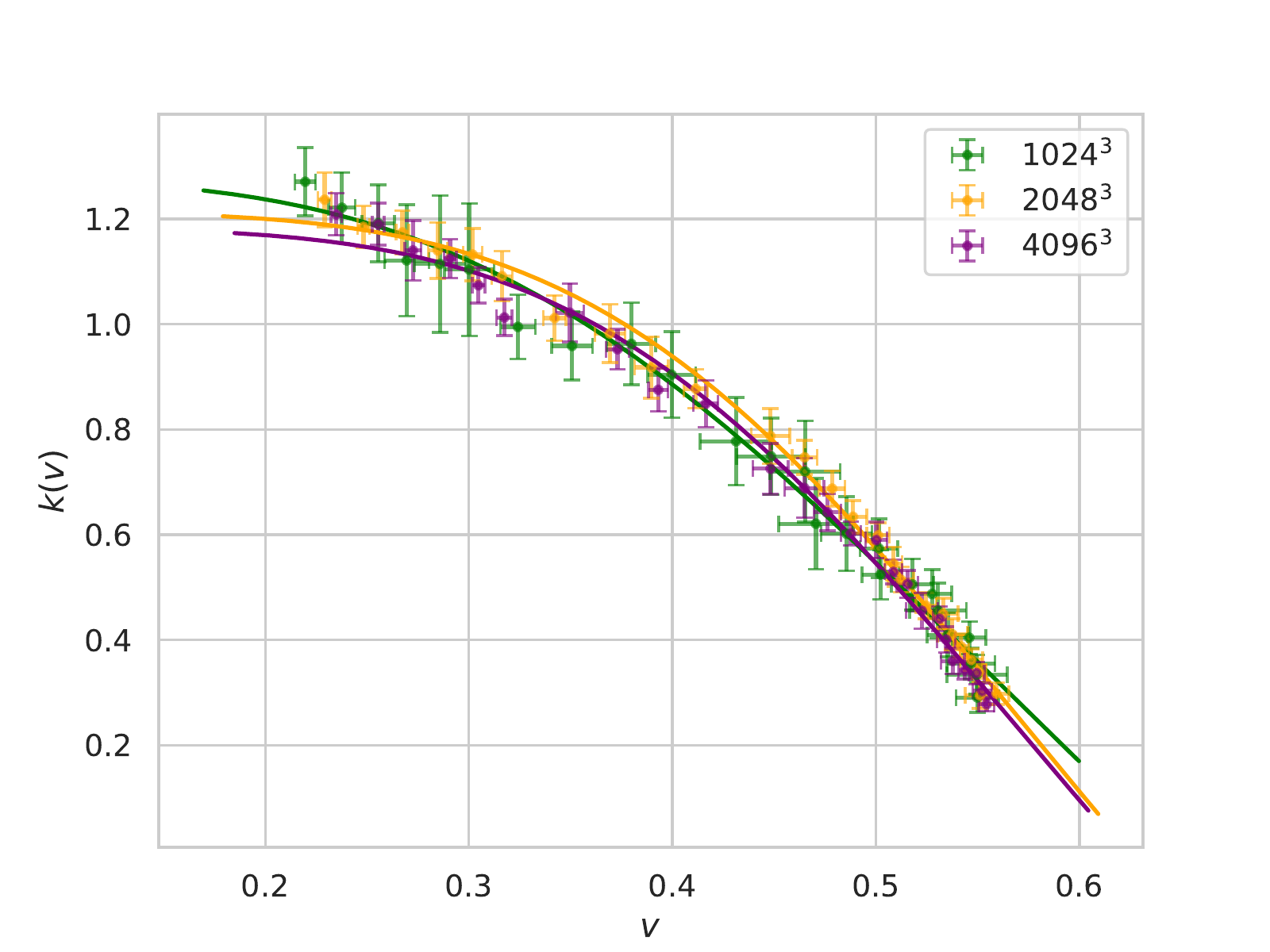}
  \includegraphics[width=1.0\columnwidth]{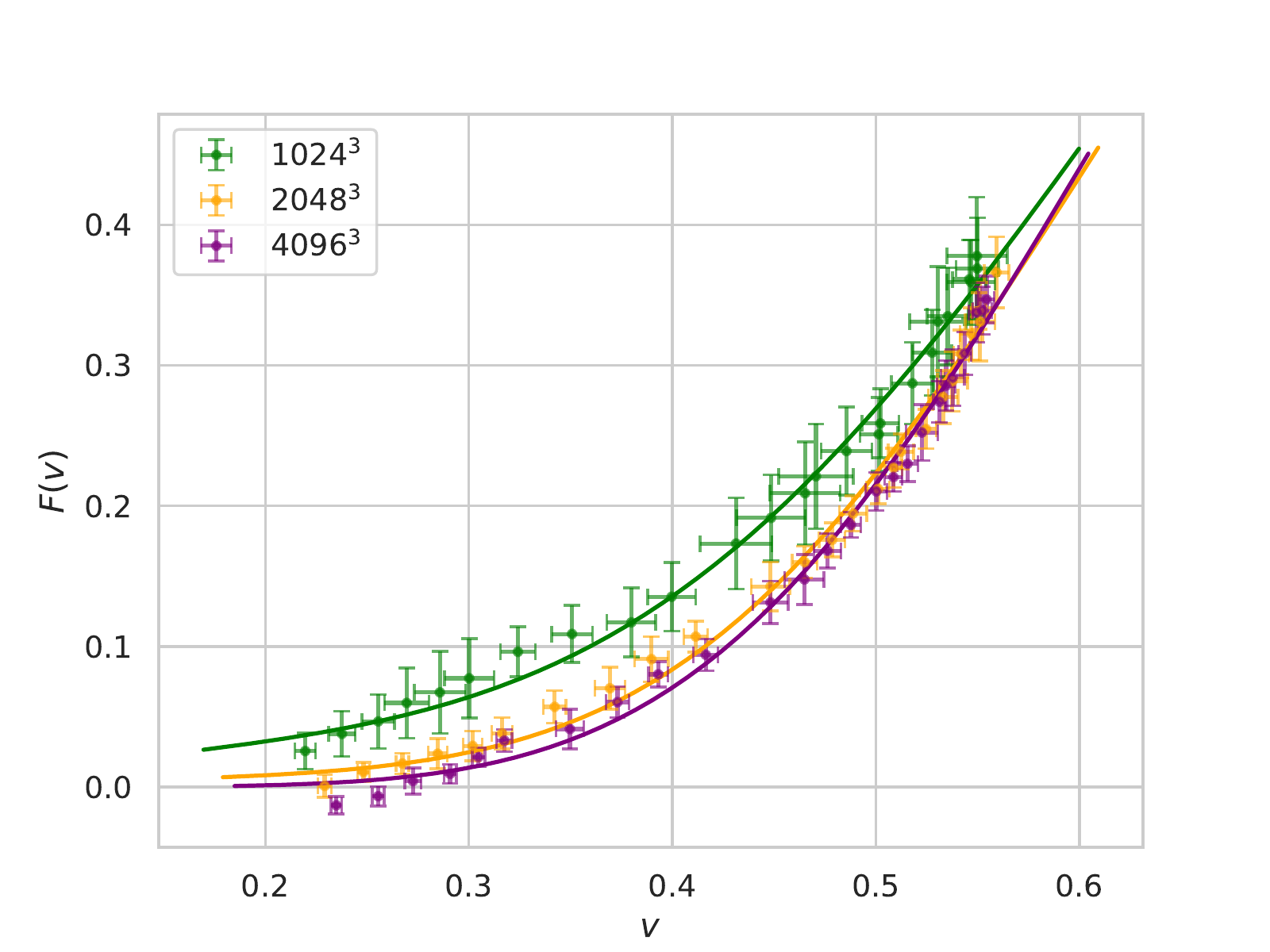}
  \caption{Comparison of the mean rate of change of correlation length $\xi / \eta$ (top left) and the mean velocity $\langle v \rangle$ (top right) with the solid lines corresponding to the calibration and the shaded regions to the uncertainty of the measurements of each estimator for three different box sizes. The bottom plots show how these differences impact the momentum parameter $k(v)$ (bottom left) and in the energy loss parameter $F(v)$ (bottom right).}
  \label{fig2}
\end{figure*}

\section{\label{cal}Improving the calibration: Dynamic range and lattice size}

In order to calibrate the VOS for different simulation sets, we will use the same Markov Chain Monte Carlo (MCMC) procedure used in \cite{Correia:2020gkj} (based on the emcee Python package \cite{emcee}). This will allow us to sample from the 6 dimensional parameter space to find posterior distributions for each parameter, and thus understand possible correlations, predict expected uncertainties and retrieve likelihood maxima. The logarithm of the likelihood is obtained via the $\chi^2$ statistic. We assume uniform distributions for all priors. In all cases we use 32 walkers and a minimum of 10000 steps. The ranges of the parameters intervals in the uniform distributions and the number of steps are different in the various cases, as is the time needed for convergence.

We now set out to understand how the dynamic range and lattice size of the simulations can affect the model calibration. We will first explore lattice sizes of $1024^3$, $2048^3$ and $4096^3$, with a common lattice spacing of $\Delta x = 0.5$, in 25 different expansion rates from $m=0.45$ to $m=0.95$) and using 10 runs per each expansion rate. We start by using the equation of state estimator for the velocities. It has been reported in the literature that there is a slow drift in the values of $\xi/\eta$ \cite{Hindmarsh:2017qff,Correia:2020yqg}. This is often partially ascribed (when going from $512^3$ to $1024^3$) to the different cooled versus non-cooled initial conditions. We have previously quantified the possible impacts of cooling on the VOS model calibration in Paper 2, and here we apply no cooling. Therefore any remaining drift should be uniquely determined by lattice size. Our results confirm this small drift. This is clearly seen in the top-left panel of figure \ref{fig2}, where we plot the calibrated VOS prediction for $\xi / \eta$ and the shaded regions correspond to the measured simulation values (including their statistical uncertainties). 

Besides the drift in $\xi / \eta$, a reduction in the uncertainties is also clearly visible, which is fully expected since the larger dynamic range lessens the impact of any systematics due to the initial conditions, including in particular any effect of the numerical offset $\eta_0$. It is interesting to note that, as has been previously found for domain walls \cite{Rybak1,Rybak2}, there is a lattice resolution beyond which the values seem to stabilize: qualitatively there is a discernible change when going from $1024^3$ to $2048^3$, but there seems to be little change when going from $2048^3$ to $4096^3$. In particular, no change is observed for the mean string velocities, as seen in the top right panel of \ref{fig2}.

\begin{figure*}
\centering
  \includegraphics[width=0.9\paperwidth]{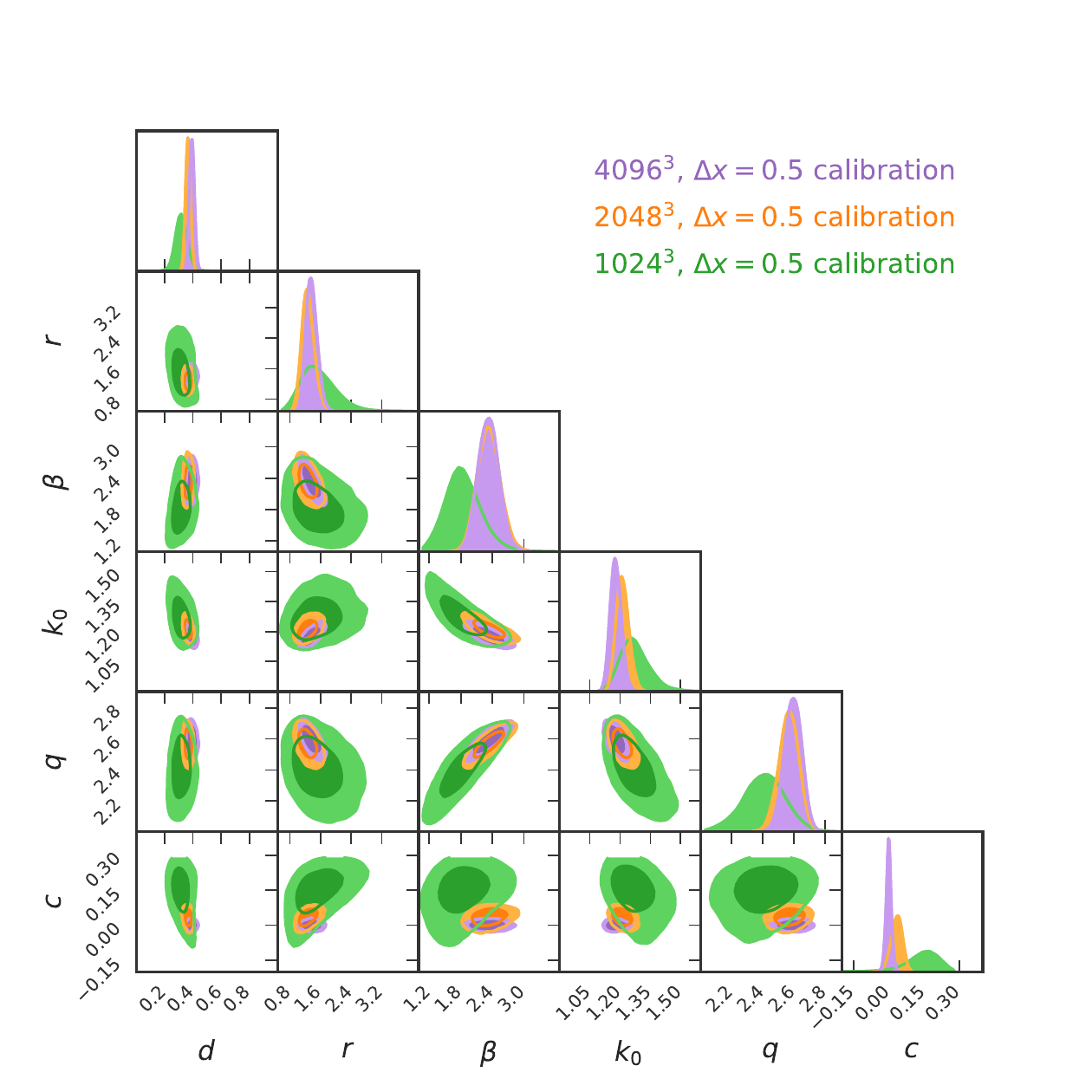}
  \caption{Corner plots for the MCMC calibration of the VOS model, obtained with the velocity estimator $\langle v_\omega \rangle$, for three different box sizes. The 2D panels the depict the $1\sigma$ and $2\sigma$ confidence regions.}
  \label{fig3}
\end{figure*}

Naturally, these differences impact the two velocity-dependent functions, as can be seen in the two bottom panels of Fig. \ref{fig2}. In particular the energy loss function $F(v)$ is affected by the drift of $\xi / \eta$ by being shifted downwards, which seems to suggest a change in the value of the loop chopping parameter $c$ and/or in the normalization of the radiation term $d$. The impact on the momentum parameter $k(v)$ is much smaller, and is limited to its maximum value being slightly reduced. We can confirm these and explore how the VOS model parameters are affected by looking at the corner plots of $1\sigma$ and $2\sigma$ contours of the model calibration, depicted in Fig. \ref{fig3}, and the corresponding parameter values in the top half of Table \ref{table1}. Visually this confirms the reduction of the values of $k_0$ and $c$. Notice the significant reduction in the area of the 2D confidence regions, which follows from the smaller uncertainties. Doubling the box size leads to the gain of a factor of 16 in statistical constraining power: a factor of eight in volume and a factor of two in dynamic range. The fact that almost all parameters seem to not change by increasing the resolution from $2048^3$ to $4096^3$ is also manifest. The only parameters which change slightly from $2048^3$ to $4096^3$ boxes are $c$ and (to a lesser extent) $d$. Both are related to the energy loss function, which explains why they are anticorrelated: as $c$ decreases, $d$ increases. This seems to suggest that as the lattice resolution increases, energy loss through loop production is gradually replaced by radiative losses, eventually becoming negligible at $4096^3$ for this choice of velocity estimator. 

\begin{table*}

\begin{tabular}{ c | c | c |  c  c  c  c  c  c }
\hline
Lattice size & $\Delta x$ & Velocity estimator & d & r & $\beta$ & $k_0$ & q & c \\
\hline
$1024^3$ &     &                              & $0.32^{+0.04}_{-0.04}$ & $1.51^{+0.48}_{-0.37}$ & $1.82^{+0.34}_{-0.30}$ & $1.27^{+0.08}_{-0.06}$ & $2.41^{+0.13}_{-0.13}$ & $0.15^{+0.05}_{-0.07}$ \\
$2048^3$ & 0.5 & $\langle v_\omega^2 \rangle$ & $0.37^{+0.02}_{-0.02}$ & $1.27^{+0.17}_{-0.15}$ & $2.33^{+0.21}_{-0.20}$ & $1.21^{+0.03}_{-0.03}$ & $2.57^{+0.06}_{-0.06}$ & $0.03^{+0.02}_{-0.03}$  \\
$4096^3$ &     &                              & $0.39^{+0.02}_{-0.02}$ & $1.36^{+0.15}_{-0.13}$ & $2.32^{+0.20}_{-0.18}$ & $1.18^{+0.03}_{-0.03}$ & $2.59^{+0.05}_{-0.05}$ & $0.00^{+0.01}_{-0.01}$  \\
\hline
$1024^3$ &     &                              & $0.35^{+0.23}_{-0.10}$ & $2.39^{+1.58}_{-0.94}$ & $2.79^{+0.73}_{-0.56}$ & $1.06^{+0.05}_{-0.05}$ & $2.95^{+0.18}_{-0.19}$ & $0.44^{+0.04}_{-0.05}$ \\
$2048^3$ & 0.5 & $\langle v_\phi^2 \rangle$   & $0.33^{+0.05}_{-0.04}$ & $1.86^{+0.39}_{-0.32}$ & $2.65^{+0.28}_{-0.26}$ & $1.05^{+0.03}_{-0.03}$ & $2.84^{+0.08}_{-0.08}$ & $0.31^{+0.02}_{-0.02}$  \\
$4096^3$ &     &                              & $0.36^{+0.03}_{-0.03}$ & $1.72^{+0.26}_{-0.23}$ & $2.50^{+0.21}_{-0.20}$ & $1.06^{+0.02}_{-0.02}$ & $2.83^{+0.06}_{-0.06}$ & $0.23^{+0.01}_{-0.01}$  \\
\hline
\end{tabular}
\caption{Calibrated VOS model parameters for our three different lattice sizes, $1024^3$, $2048^3$ and $4096^3$, all with the same lattice spacing $\Delta x=0.5$, and two different choices of velocity estimators, $\langle v^2_\omega \rangle$ and $\langle v^2_\phi \rangle$ (in the top and bottom parts of the table, respectively), further described in the main text. Displayed values correspond to 16th, 50th, 84th percentiles of the posterior distributions. \label{table1}}
\end{table*}

\begin{figure*}
\centering
  \includegraphics[width=0.9\paperwidth]{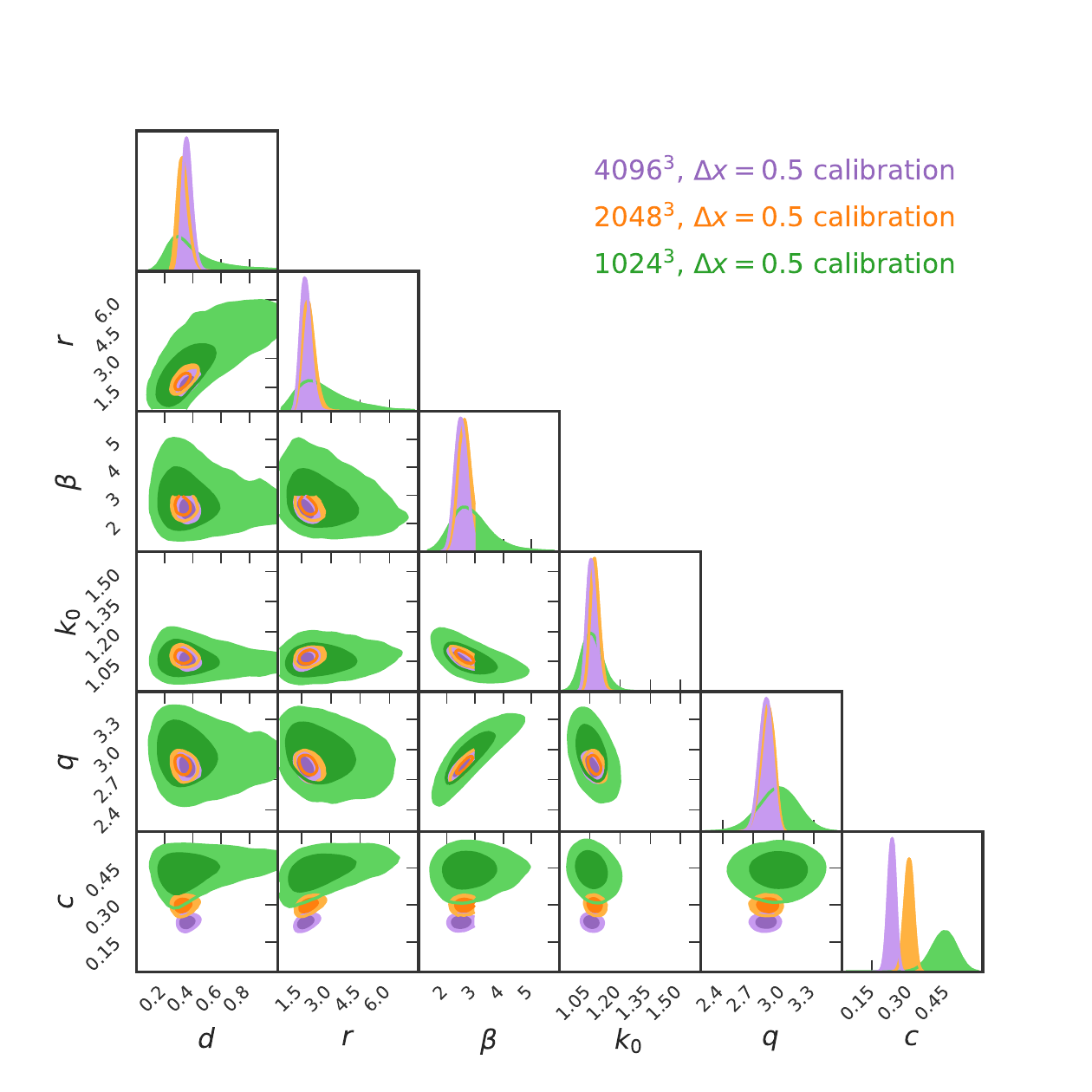}
  \caption{Corner plots for the MCMC calibration of the VOS model, obtained with the velocity estimator $\langle v_\phi \rangle$, for three different box sizes. The 2D panels the depict the $1\sigma$ and $2\sigma$ confidence regions.}
  \label{fig4}
\end{figure*}

As a cross-check one can repeat the analysis replacing the equation of state estimator for the velocities by the conjugate momentum based velocity estimator. The analogous corner plots can be found in Fig. \ref{fig4} and the calibrated model parameters are in the bottom half of Table \ref{table1}. Clearly this leads to a very different calibration, and various model parameters take different values in both cases. The most notable difference is in the value of $c$: while it is still the case that increasing the lattice size reduces the value $c$, it now converges to a value that is very clearly (in a statistical sense) non-zero and comparable to that of $d$. On the other hand, in this case there is no statistically significantly drift in the value of $d$ which, somewhat remarkably, is the least affected of the six parameters.

This raises the question of the reliability and possible biases of the two velocity estimators. We note that visual inspection of the evolution of the network unambiguously reveals the formation of loops with a range of sizes (large and small), which subsequently decay. Two examples, from a $4096^3$ simulation in the radiation era, with the evolution displayed for conformal times $\eta \in [741, 1024]$, can be found in \cite{Movie1,Movie2}. The fact that $c\longrightarrow0$ if one uses the equation of state estimator for the velocities can be reconciled with this observation if this velocity estimator is not sufficiently accurate: as the statistical uncertainties in the simulation are reduced (by simulating larger boxes), this systematic uncertainty ends up dominating the analysis. We explore this issue, together with the related one of lattice spacing, in the following section.


\section{\label{lat}Lattice spacing and velocity estimators}

Although the extended VOS model generically has the six parameters introduced in Sect. \ref{VOS}, in the limit of low velocities the model reduces, to first order, to the following
\begin{equation}
  \label{eq:vos1_high}
  \frac{d\xi}{d\eta} = \frac{m \xi}{(1-m)\eta} v^2 + cv
  \end{equation}
  \begin{equation}
  \label{eq:vos2_high}
  \frac{dv}{d\eta} = (1-v^2) \bigg[ \frac{k_0}{\xi}  - \frac{2m v}{(1-m)\eta} \bigg]\,,
\end{equation}
which depends only on $c$ and $k_0$. Numerically, the low velocity limit corresponds to vary fast expansion rates, i.e. large values of $m$. The fact that these parameters are affected by the choice of velocity estimator is therefore not surprising, since our previous work in Paper 1 shows that the two velocity estimators themselves differ maximally in this limit. Specifically, the relative difference between the velocities obtained from the two estimators ranged between $6\%$ at moderate expansion rates (including the radiation era, $m=1/2$, and the matter era, $m=2/3$) and $12\%$ at $m=0.95$, which is the fastest expansion rate simulated in Paper 1.

Determining these two parameters accurately can therefore depend on how well the numerical simulation algorithm (including the estimators used for the model calibration) behaves at large expansion rates. Given the importance of $k_0$ and $c$ for understanding the amount of small-scale structure on the strings, as well as the importance of loop formation to the overall energy losses, we now turn our attention to studying the high-expansion rate limit.

We note that there are potential sources of systematics which are specific to such low velocities and can plausibly impact the VOS model calibration. For example, at highly non-relativistic velocities ($v<0.2$) it is possible that strings cannot overcome the small potential barrier present between lattice sites, known as the the Peierls-Nabarro barrier \cite{Peierls,Nabarro,Ward}. In such a case the strings become pinned at lattice sites, and do not move according to what would be physically predicted. One might then expect that this will manifest itself on the measured values of $F(v)$ and $k(v)$ as a lattice spacing dependency. 

The question, then, is: which estimator should be trusted to yield the correct velocities, and therefore the most accurate calibration of the VOS model? In \cite{Hindmarsh:2017qff} it was shown that the disagreement between velocity estimators decreases for an oscillating string in Minkoswski space as one reduces the lattice spacing. While the same reference argued that the equation of state estimator should be more reliable in Minkoswski space, this does not imply that the same applies to expanding universes, and this caveat is particularly applicable to the high expansion rates---which are of interest for the present study and are, effectively, the opposite limit to that of Minkowski space.

To better understand these systematics we begin by characterizing the differences between the two velocity estimators for large expansion rates,  specifically from $m=0.93$  to $m=0.997$. The latter choice of $m$ is as deep into the non-relativistic limit as our numerical simulation algorithm allows. Moreover, we will do this for two different sets of simulations with different lattice spacings: the standard $\Delta x = 0.5$ and what we will call the half-lattice spacing of $\Delta x =0.25$. As in the previous section, statistical uncertainties are obtained for the averages over the 10 different simulations done for each each choice of expansion rate.

\begin{figure*}
\centering
  \includegraphics[width=1.0\columnwidth]{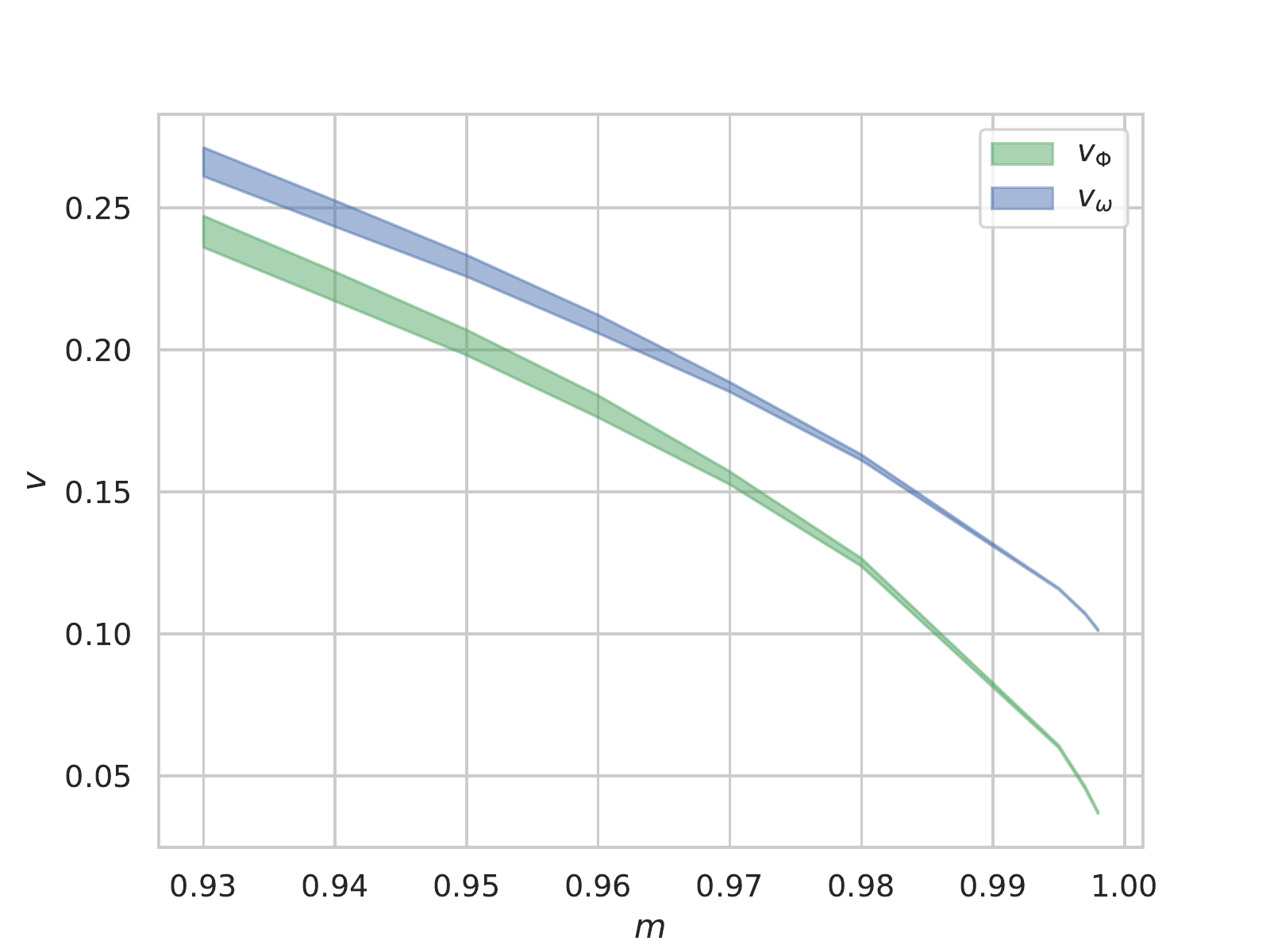}
  \includegraphics[width=1.0\columnwidth]{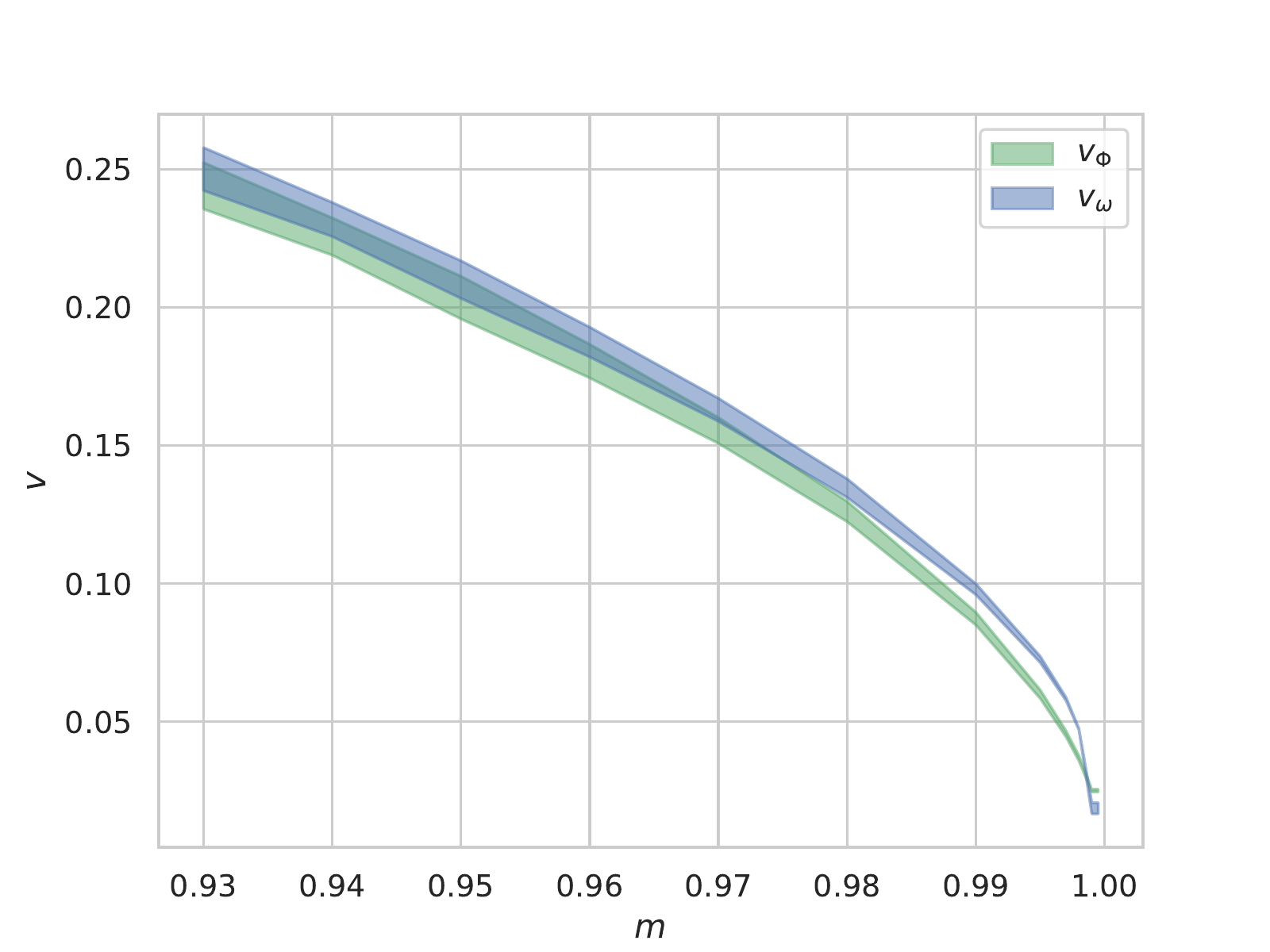}
  \includegraphics[width=1.0\columnwidth]{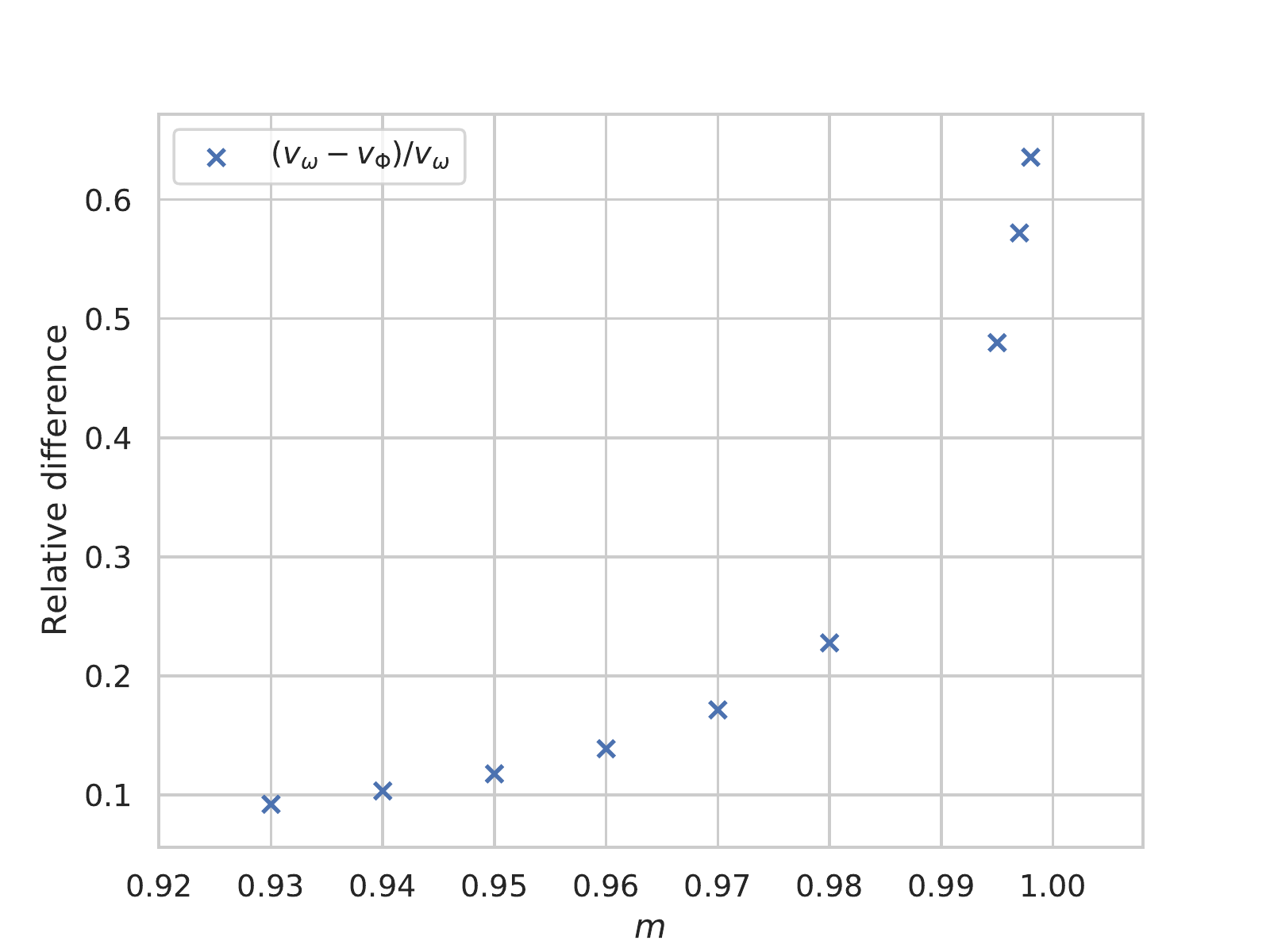}
  \includegraphics[width=1.0\columnwidth]{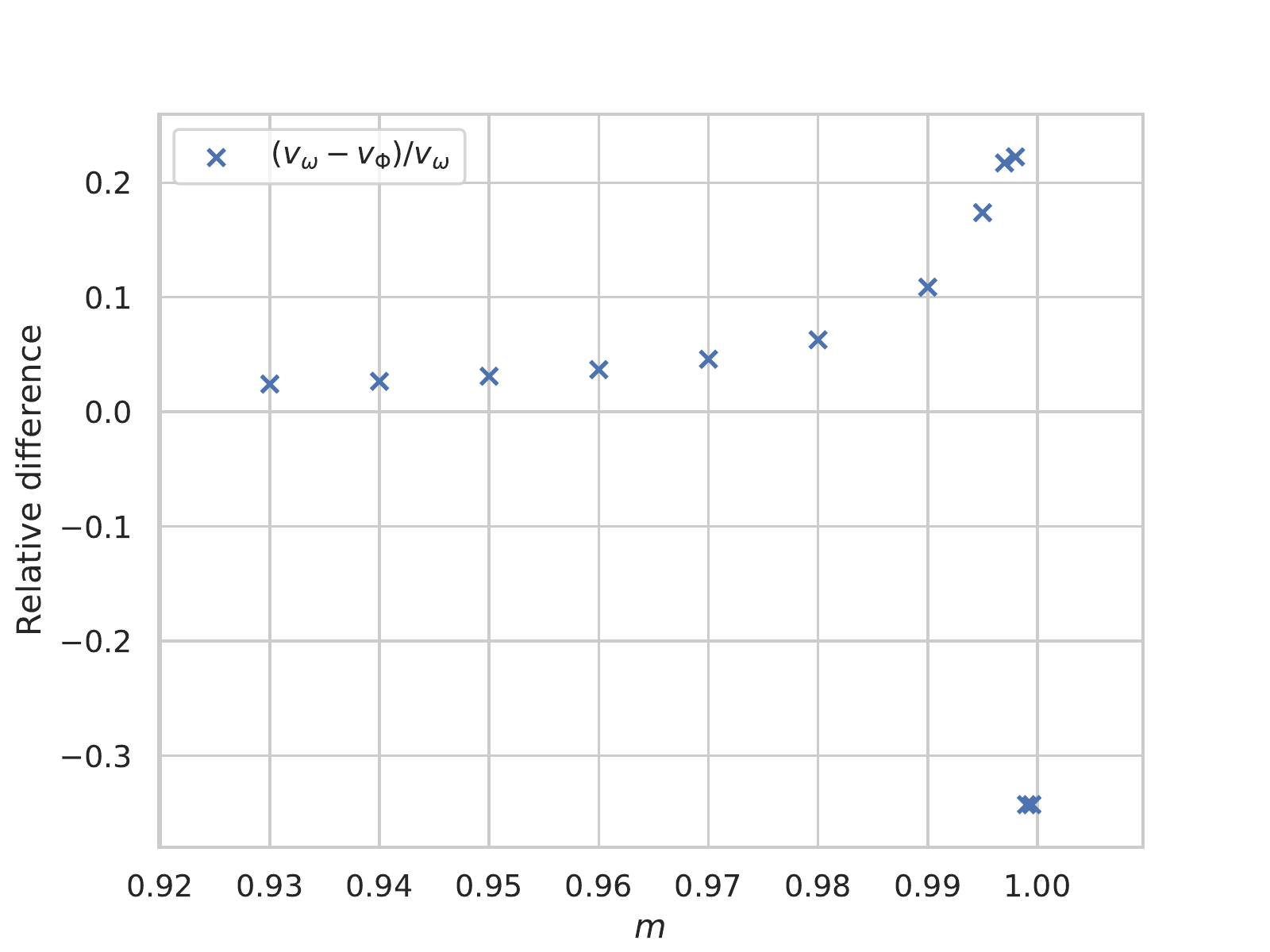}
  \caption{The effect of lattice spacing on the velocity estimators, for high expansion rate simulations. The top panels show the separate values of the velocities (with the corresponding statistical uncertainties) obtained with the two velocity estimators defined in the text, while the bottom panels show the relative difference between the two. Left and right side panels depict the results for standard spacing and half-lattice spacing.}
  \label{fig5}
\end{figure*}

\begin{figure*}
\centering
  \includegraphics[width=1.0\columnwidth]{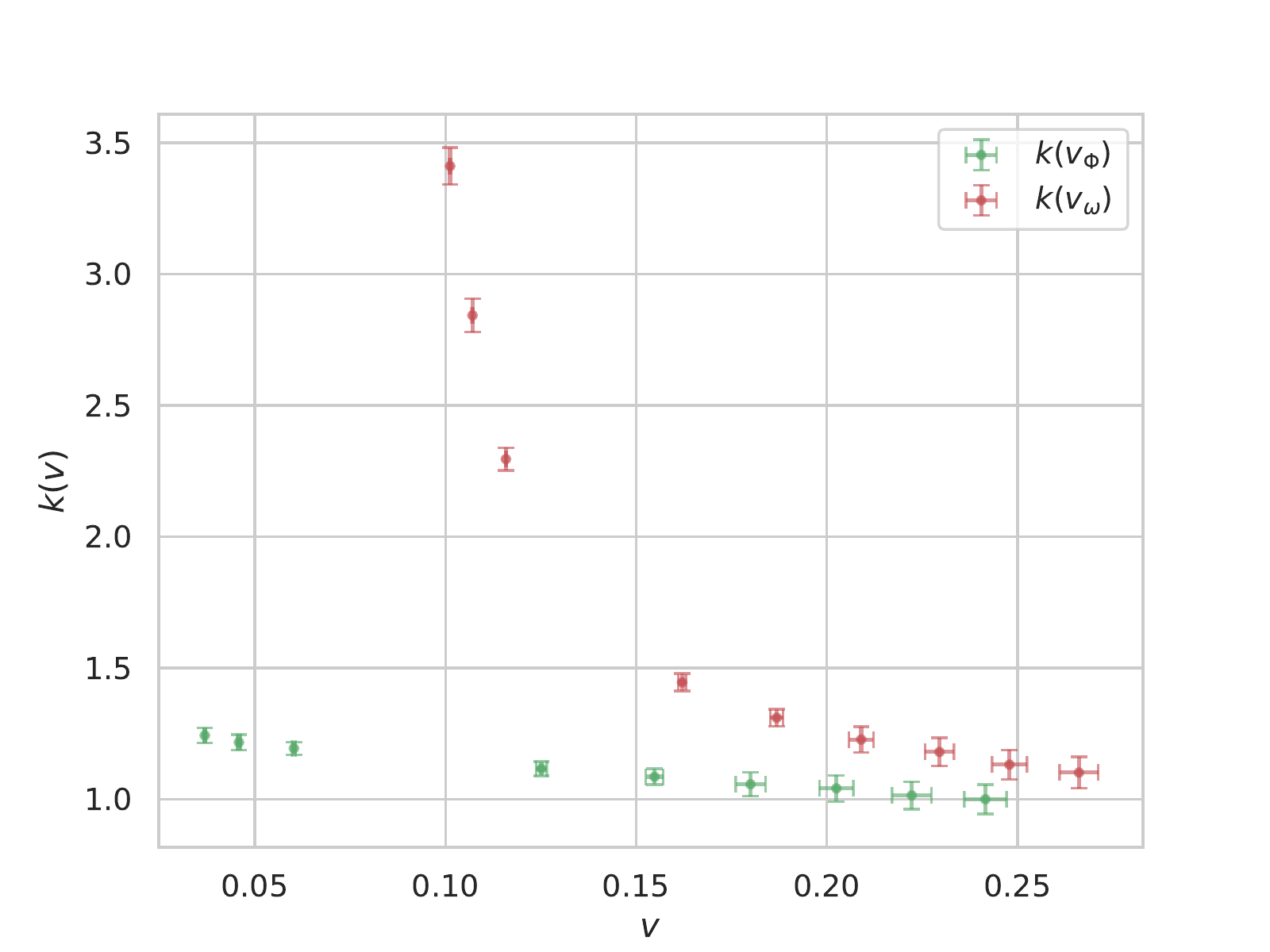}
  \includegraphics[width=1.0\columnwidth]{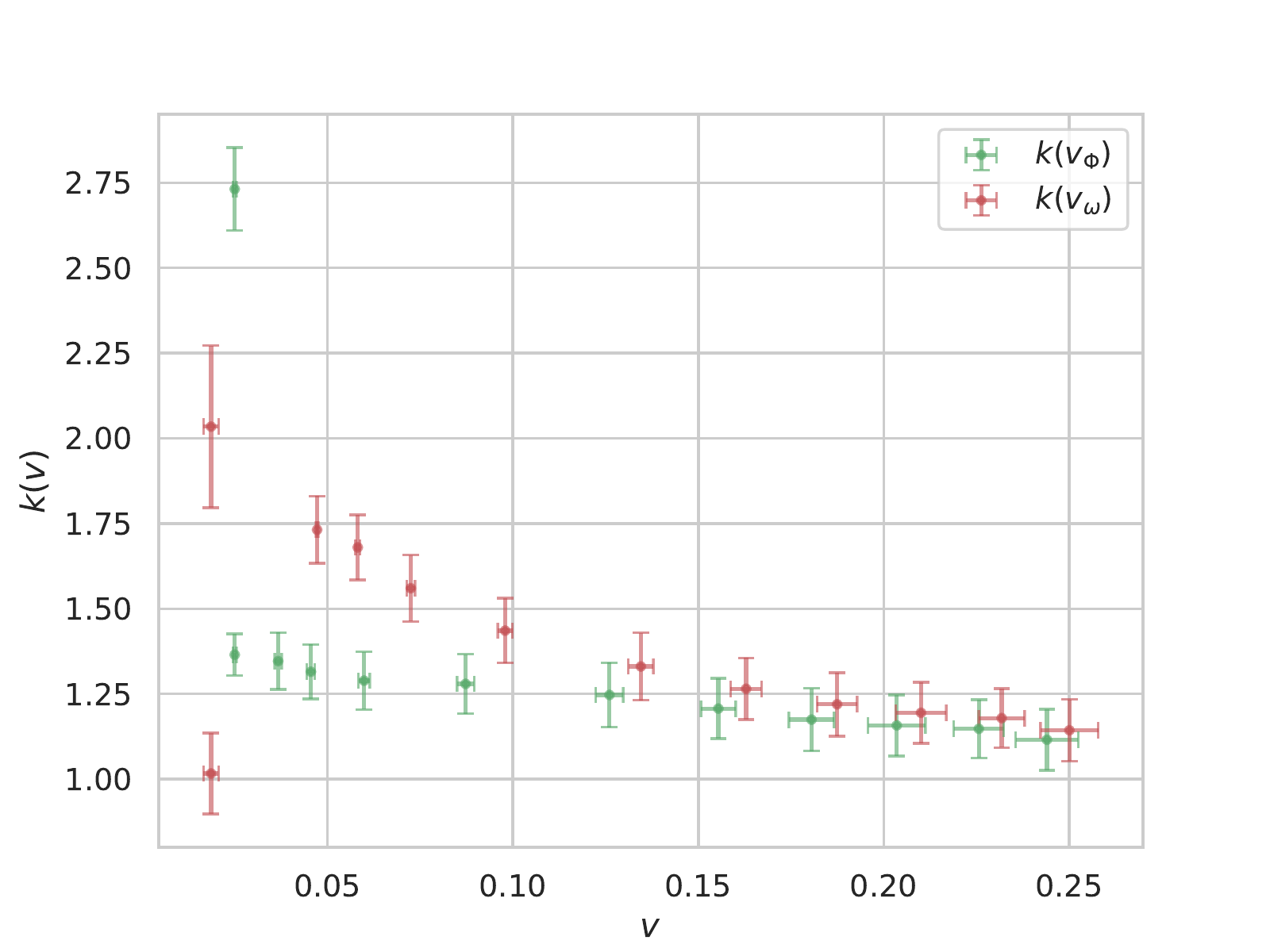}
  \includegraphics[width=1.0\columnwidth]{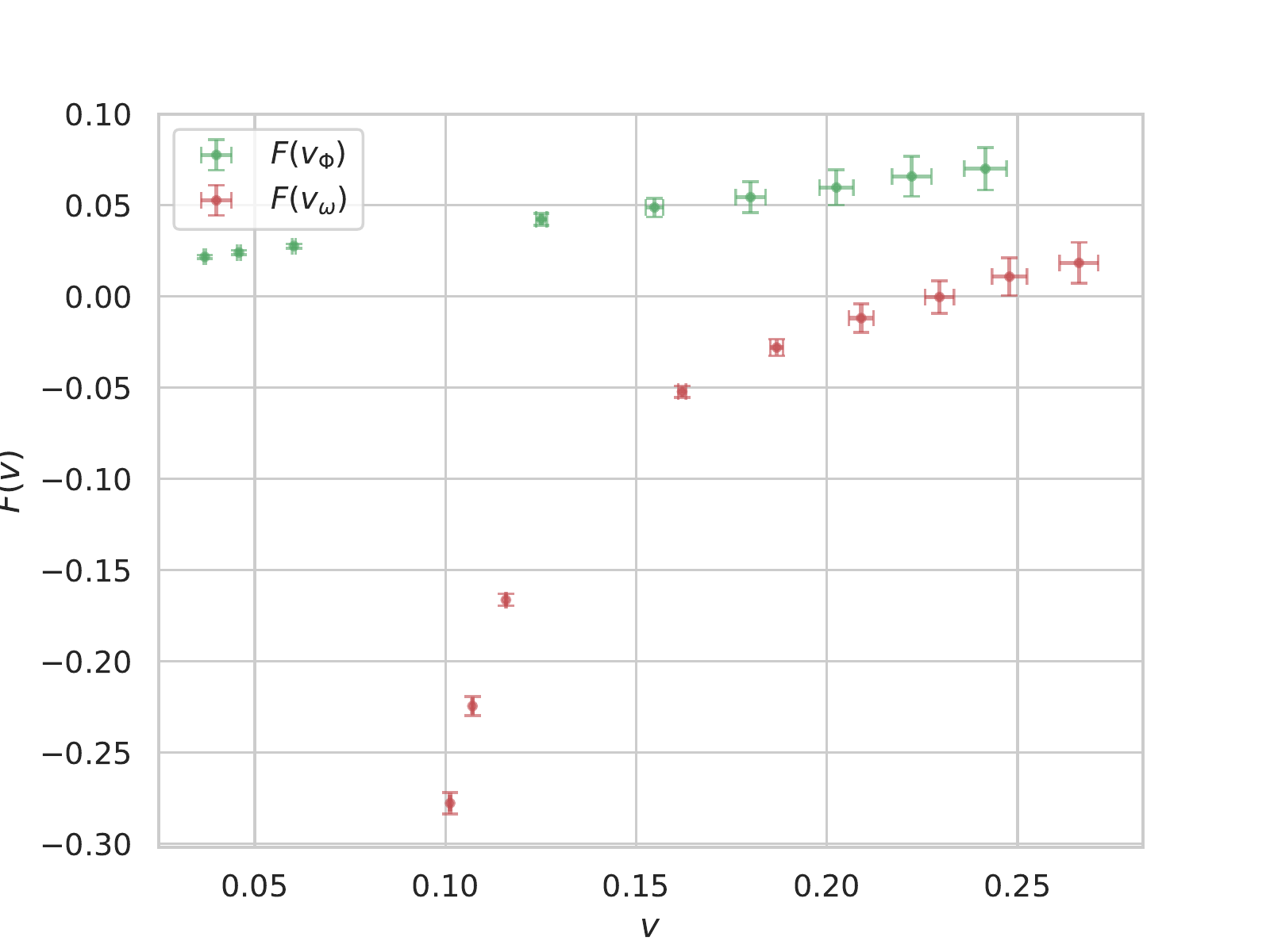}
  \includegraphics[width=1.0\columnwidth]{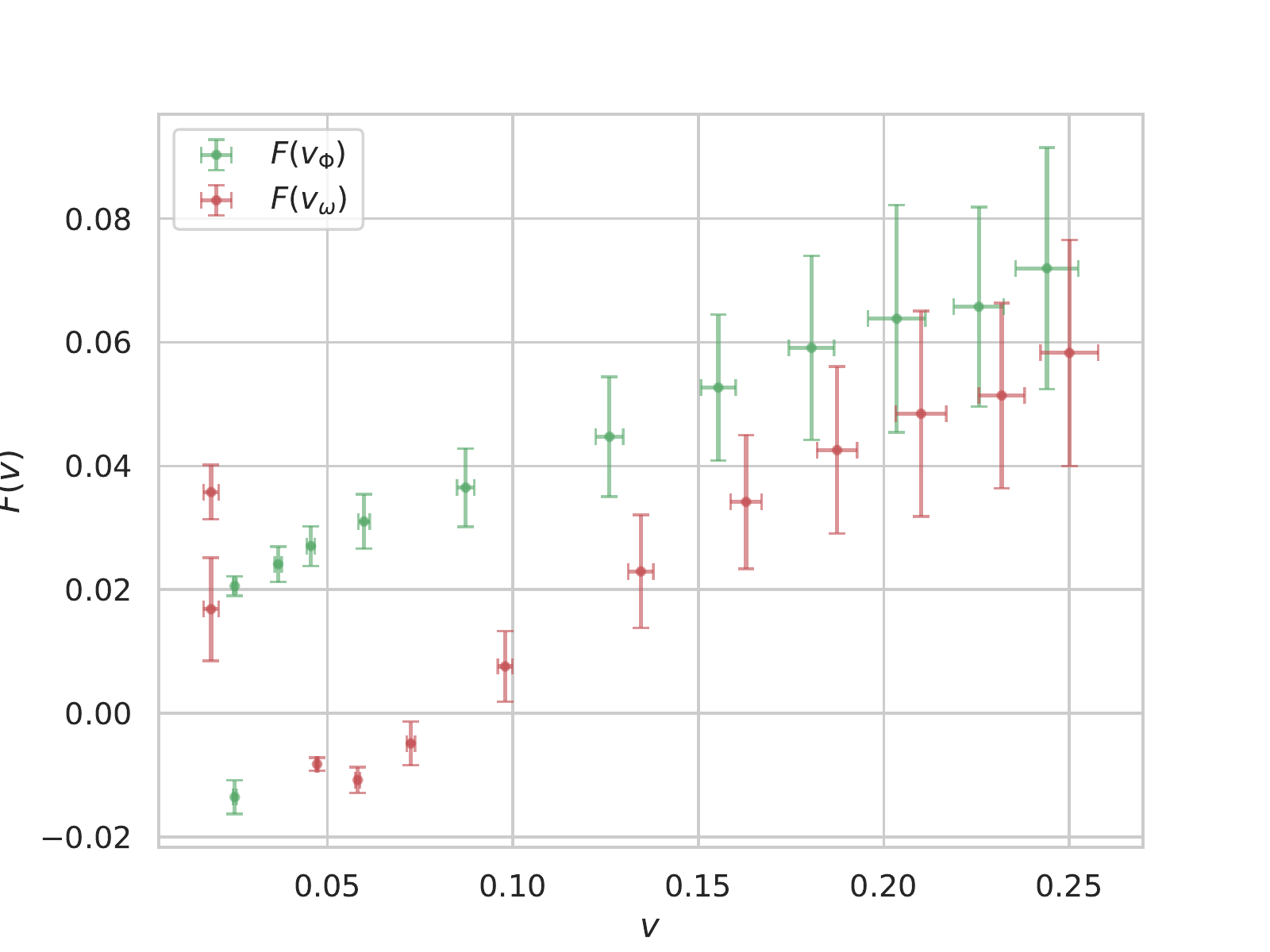}
  \caption{The effect of lattice spacing on the velocity estimators, as manifest in the velocity-dependent functions of the VOS model, for high expansion rate simulations. The top panels show the momentum parameter $k(v)$ while the bottom panels show the energy loss parameter $F(v)$, all with the corresponding statistical uncertainties, obtained with the two velocity estimators defined in the text. Left and right side panels depict the results for standard and half-lattice spacing.}
  \label{fig6}
\end{figure*}

The velocities obtained from both estimators are shown in Fig. \ref{fig5}. Manifestly the difference between the two estimators increases with the expansion rate, but equally clearly it decreases with the lattice spacing. Thus we confirm that a small lattice spacing is crucial for an accurate calibration. As before, these differences are reflected in the inferred values of $F(v)$ and $k(v)$ which can be seen in Fig. \ref{fig6}. It is worthy of note that in standard lattice spacing the equation of state estimator mostly fails to give a physically reasonable (positive) value for almost all the large expansion rates considered, casting doubt into how reliable it can be, even beyond these expansion rates. When the lattice size is decreased, the disagreement decreases as well, but the equation of state estimator can still lead to negative values for the energy loss parameter. At the highest expansion rates that we have simulated, even our half-lattice spacing cannot completely remove the disagreement between the tow estimators. Overall, it can also be seen that the estimator that changes the most with a reduction of the lattice spacing is the equation of state one, which further supports the interpretation that it is the least trustworthy of the two, at least in the high expansion rate limit. Note that this does not contradict the claim, in \cite{Hindmarsh:2017qff}, that it is the more accurate one in the opposite limit of Minkowski space. As has already been pointed out above, the fast expansion and Minkowski are effectively the two opposite ends of the spectrum, but there are several other examples in the dynamics of cosmic string networks which show that Minkowski space may not be representative of the evolution in expanding universes \cite{Fractal}.

Returning to our calibration in the relativistic range of velocities, the three expansion rates $m=0.93, 0.94, 0.95$ give reasonably similar predictions for the energy loss slope and momentum parameter. Having then established which estimator is more reliable at high expansion rates, we should understand the impact of the estimator choice in the opposite limit of (comparatively) low expansion rates and high velocities. Towards this goal we can perform an additional check, namely to compare the calibration at $4096^3$ resolution and half-lattice spacing with the one in the previous section at $2048^3$ with standard spacing. Note that these two correlated choices of lattice size and lattice spacing ensures a fair comparison, since the two sets of simulations will have the same dynamic range. 

\begin{figure*}
\centering
  \includegraphics[width=0.9\paperwidth]{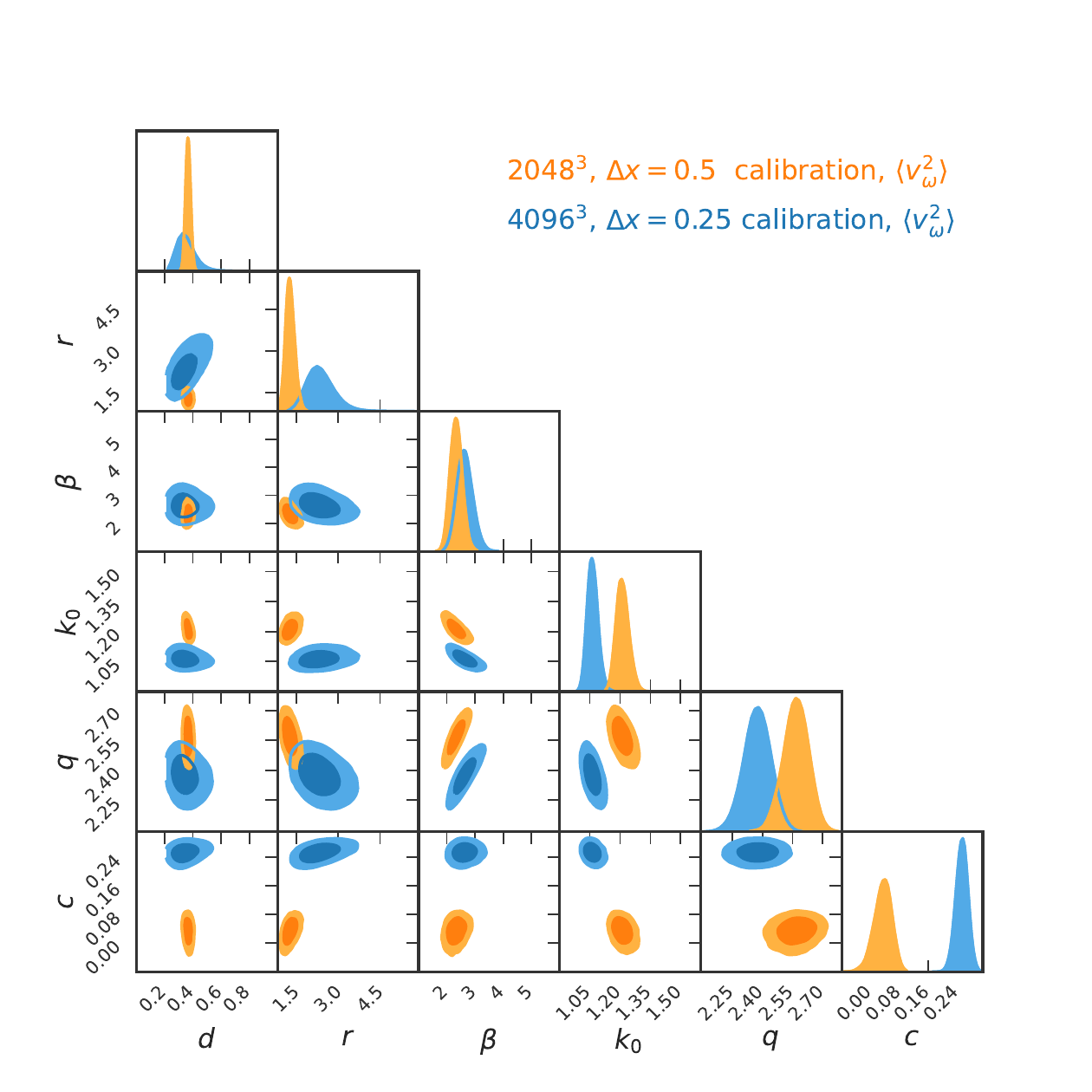}
  \caption{Corner plots for the MCMC calibration of the VOS model, obtained with the velocity estimator $\langle v_\omega \rangle$ and two different lattice spacings $\Delta x = 0.5$ and $\Delta x = 0.25$. The 2D panels the depict the $1\sigma$ and $2\sigma$ confidence regions.}
  \label{fig7}
\end{figure*}

\begin{figure*}
\centering
  \includegraphics[width=0.9\paperwidth]{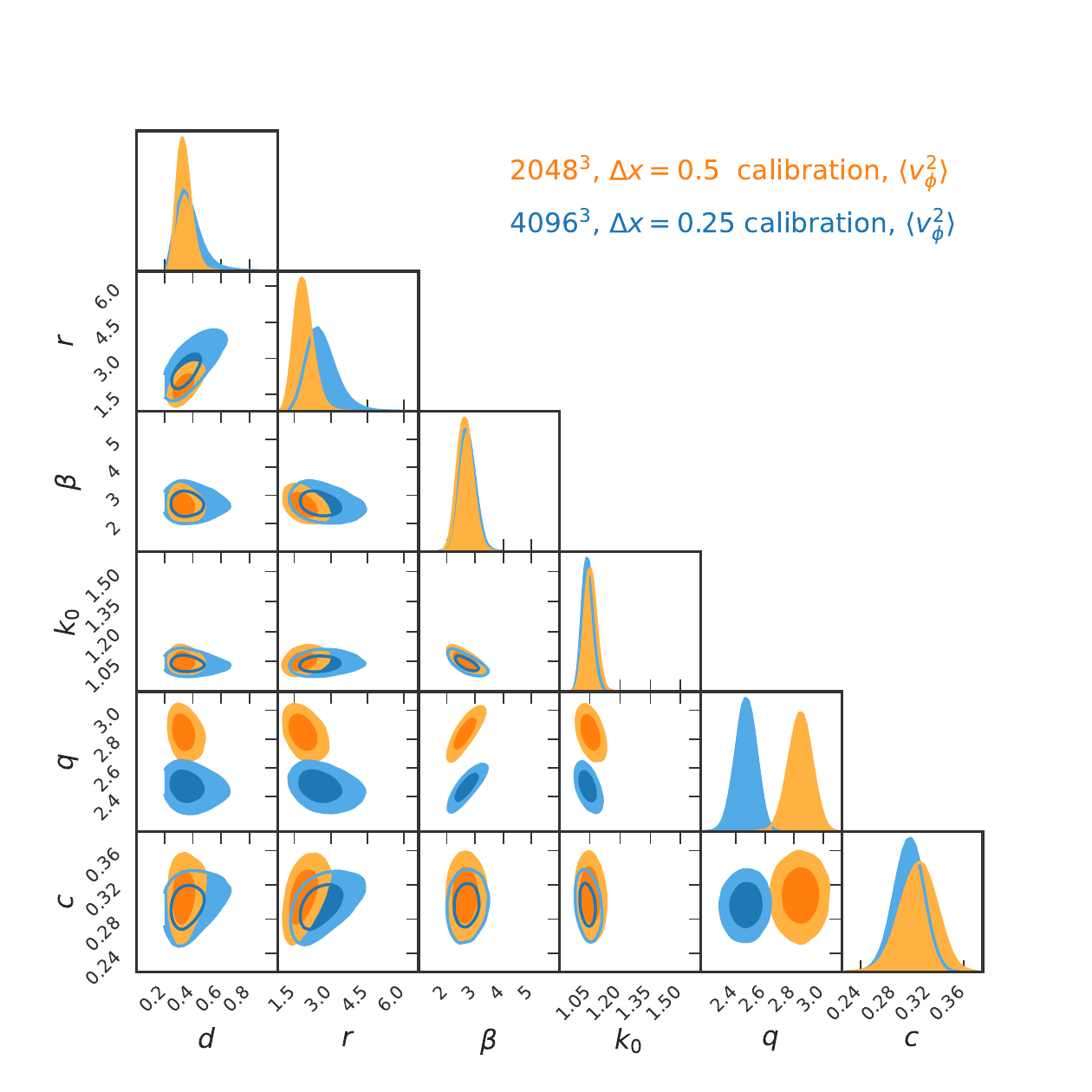}
  \caption{Corner plots for the MCMC calibration of the VOS model, obtained with the velocity estimator $\langle v_\phi \rangle$ and two different lattice spacings $\Delta x = 0.5$ and $\Delta x = 0.25$. The 2D panels the depict the $1\sigma$ and $2\sigma$ confidence regions.}
  \label{fig8}
\end{figure*}

The resulting calibration can be seen in Figs. \ref{fig7} and \ref{fig8}, as well as in Table \ref{table2}. Here we see something interesting. For the equation of state estimator the calibration changes drastically, with 4 of the 6 VOS model parameters (including the loop chopping efficiency $c$) being changed by several standard deviations; the only parameters that are unchanged (within their statistical uncertainties) are $d$ and $\beta$. For the conjugate momentum estimator the result is almost the opposite: the calibration is far more stable and only the VOS model parameter $q$ is significantly affected by the choice of lattice spacing.

\begin{table*}

\begin{tabular}{ c | c | c |  c  c  c  c  c  c }
\hline
Lattice size & $\Delta x$ & Velocity estimator & d & r & $\beta$ & $k_0$ & q & c \\
\hline
$2048^3$ & 0.5 & $\langle v_\omega^2 \rangle$ & $0.37^{+0.02}_{-0.02}$ & $1.27^{+0.17}_{-0.15}$ & $2.33^{+0.21}_{-0.20}$ & $1.21^{+0.03}_{-0.03}$ & $2.57^{+0.06}_{-0.06}$ & $0.03^{+0.02}_{-0.03}$  \\
$4096^3$ & 0.25&                              & $0.34^{+0.07}_{-0.05}$ & $2.32^{+0.52}_{-0.40}$ & $2.62^{+0.29}_{-0.26}$ & $1.06^{+0.03}_{-0.02}$ & $2.37^{+0.06}_{-0.07}$ & $0.25^{+0.02}_{-0.02}$  \\
\hline
$2048^3$ & 0.5 & $\langle v_\phi^2 \rangle$   & $0.33^{+0.05}_{-0.04}$ & $1.86^{+0.39}_{-0.32}$ & $2.65^{+0.28}_{-0.26}$ & $1.05^{+0.03}_{-0.03}$ & $2.84^{+0.08}_{-0.08}$ & $0.31^{+0.02}_{-0.02}$  \\
$4096^3$ & 0.25&                              & $0.36^{+0.09}_{-0.06}$ & $2.56^{+0.64}_{-0.50}$ & $2.69^{+0.30}_{-0.27}$ & $1.04^{+0.03}_{-0.02}$ & $2.47^{+0.07}_{-0.07}$ & $0.30^{+0.02}_{-0.02}$  \\
\hline
\end{tabular}
\caption{Calibrated VOS model parameters for our two choices of lattice spacing $\Delta x$ and corresponding lattice sizes, for the two different choices of velocity estimators, $\langle v^2_\omega \rangle$ and $\langle v^2_\phi \rangle$, further described in the main text. Displayed values correspond to 16th, 50th, 84th percentiles of the posterior distributions. \label{table2}}
\end{table*}

Therefore, and despite the existence of degeneracies between the VOS model parameters (clearly visible in our the four sets of corner plots), our analysis strongly suggests that the conjugate momentum estimator for velocities is physically more reliable for expanding universes (or, at the very least, for relatively fast expansion rates), notwithstanding the fact that in Minkowski space the opposite result may hold  \cite{Hindmarsh:2017qff}.
 
\section{\label{cls}Coda: Observational impact of different calibrations}

In the previous sections we have provided a new calibration for the VOS model parameters, which is qualitatively similar to, but quantitatively different from, those of Paper 1 and Paper 2. Additionally, it also differs from the calibration of the previous version of the VOS model \cite{Book}. The question therefore arises as to the impact of these different calibrations on observational constraints on cosmic strings. While a detailed study of this issue is beyond the scope of the present work, in this section we will nevertheless provide a brief illustration of this impact.

Towards this end, we will compute the expected anisotropies in the Cosmic Microwave Background due to cosmic string networks, under different VOS model calibrations, using the publicly available code CMBACT4 \cite{Pogosian:1999np}. This code is based on the so-called unconnected segments model \cite{Albrecht,Battye}, which is based on several simplistic assumptions and and is understood to be no better than order of magnitude accurate. More robust methods for calculating these anisotropies, directly from simulations, are known both for field theory simulations \cite{Bevis:2006mj,PlanckDefects} and for Nambu-Goto simulations \cite{PlanckDefects,Lazanu1,Lazanu2}. Still, our goal here is merely to illustrate the relative differences which stem from the different calibrations, leaving a detailed comparison with observations for future work.

Specifically, we will consider four different calibrations. The first two rely on the the standard VOS calibration, without explicit radiative energy loss terms, and a $k(v)$ function inferred from Nambu-Goto simulations as described in \cite{Martins:2000cs}. These includes separate calibrations for Abelian-Higgs and Nambu-Goto  simulations, as used in for the Planck 2013 constraints \cite{PlanckDefects}; here the loop chopping parameter is set to either $c=0.57$ or $c=0.23 \pm 0.04$ respectively. The other two calibrations are for the extended VOS model, as discussed in the previous sections, for $1024^3$ boxes with $\Delta x=0.5$ lattice spacing and the equation of state velocity estimator, and for $4096^3$ boxes with $\Delta x=0.25$ lattice spacing and the conjugate momentum velocity estimator.

In each of the four cases we use 200 realizations of the CMBACT4 code, which has been previously shown to produce spectra that are as accurate as the approximations of the method allow \cite{Charnock:2016nzm}. There is an additional toy model parameter, also discussed in \cite{Charnock:2016nzm}, which can have an impact. This parameter is the so called string decay constant, $0 \leq L_f \leq 1$, which controls if, past some specific lifetime, a given string segment will cease to contribute to the overall power spectrum; in practical terms, the danger is $L_f<1$ implies that strings can start to decay earlier than their respective epoch, thus lowering the number density. Given the illustrative nature of our comparison, we keep to the standard value of $L_f = 0.5$ used by the code for all computations.

\begin{figure*}
\centering
  \includegraphics[width=1.0\columnwidth]{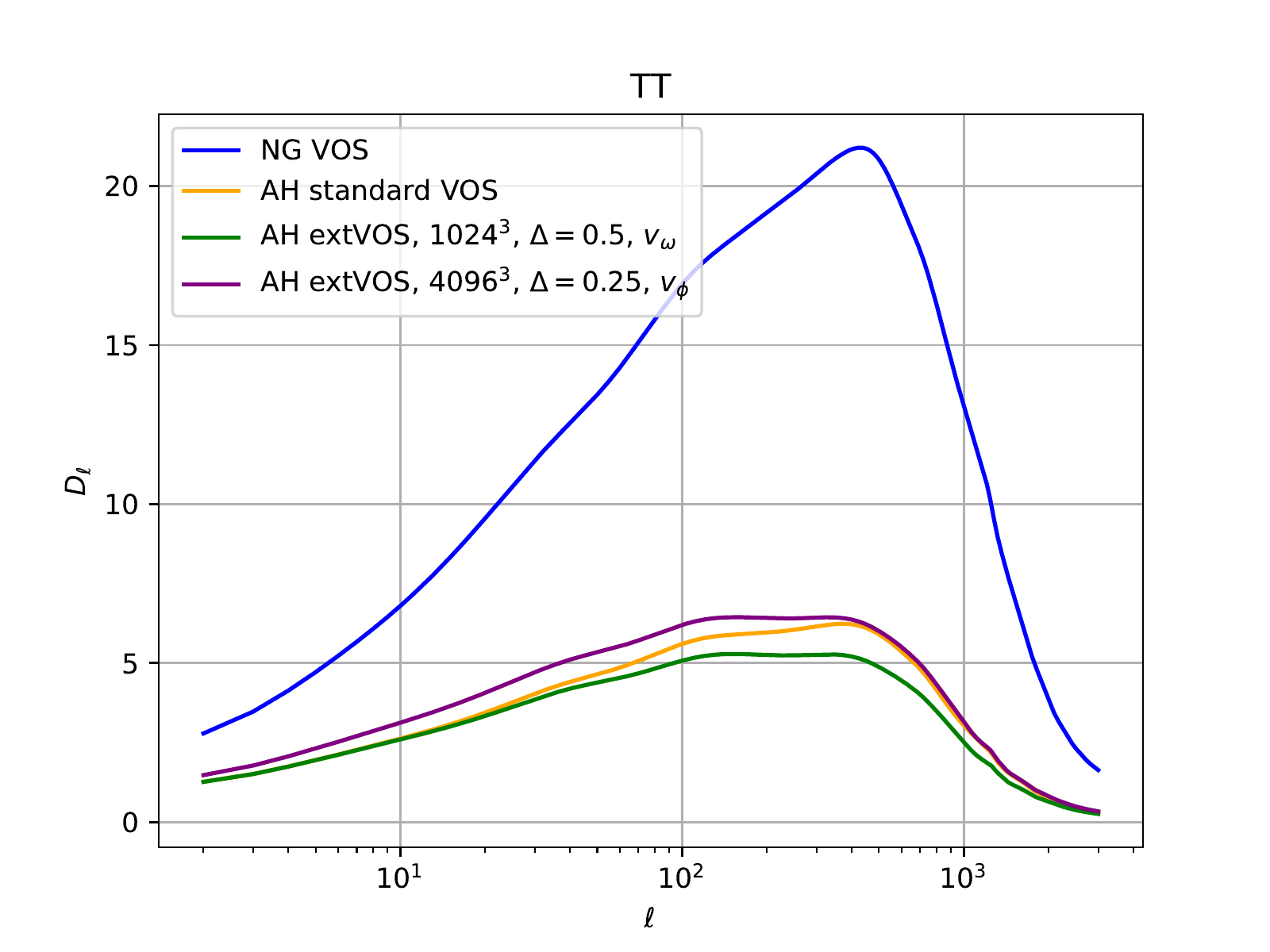}
  \includegraphics[width=1.0\columnwidth]{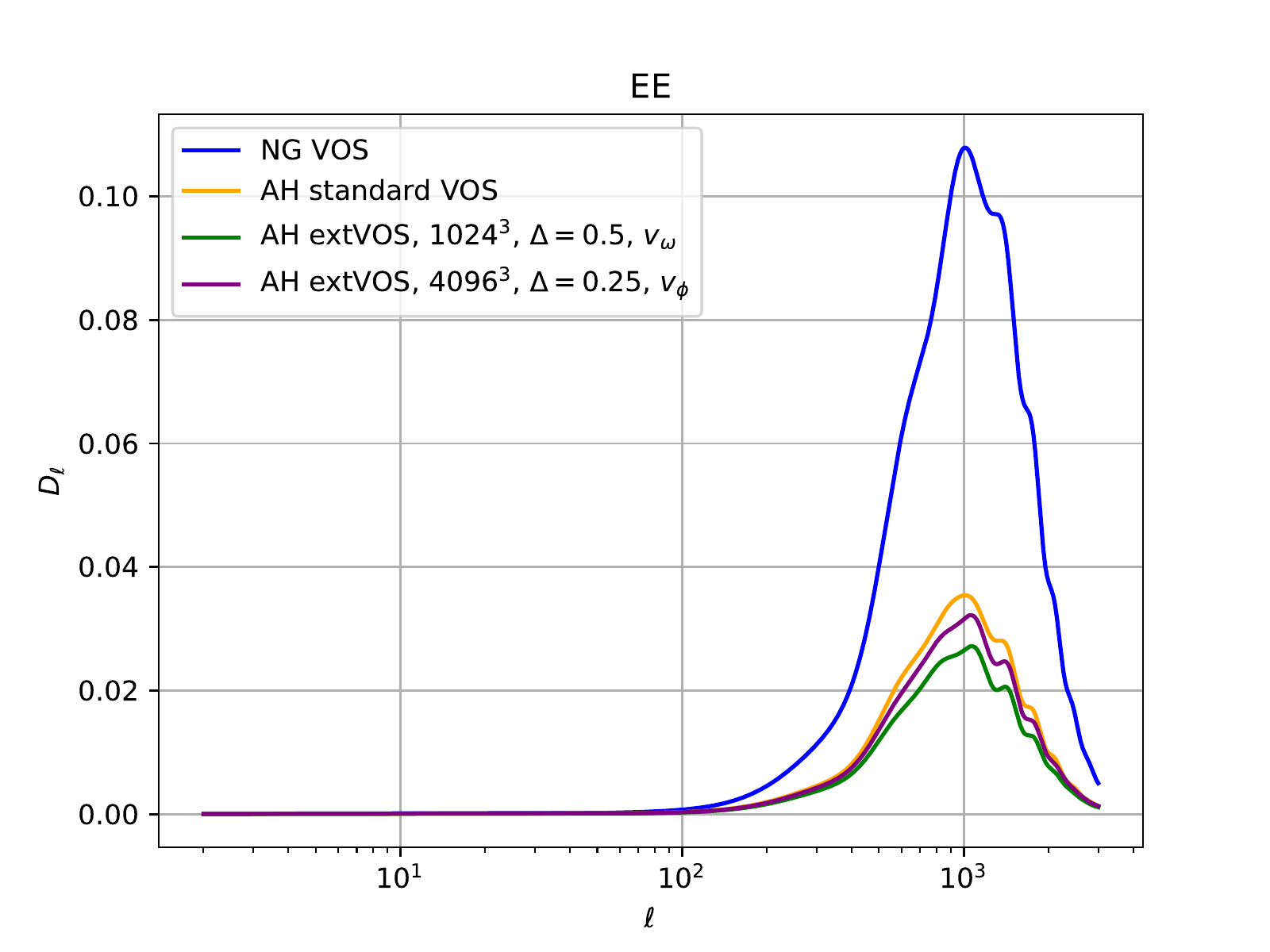}
  \includegraphics[width=1.0\columnwidth]{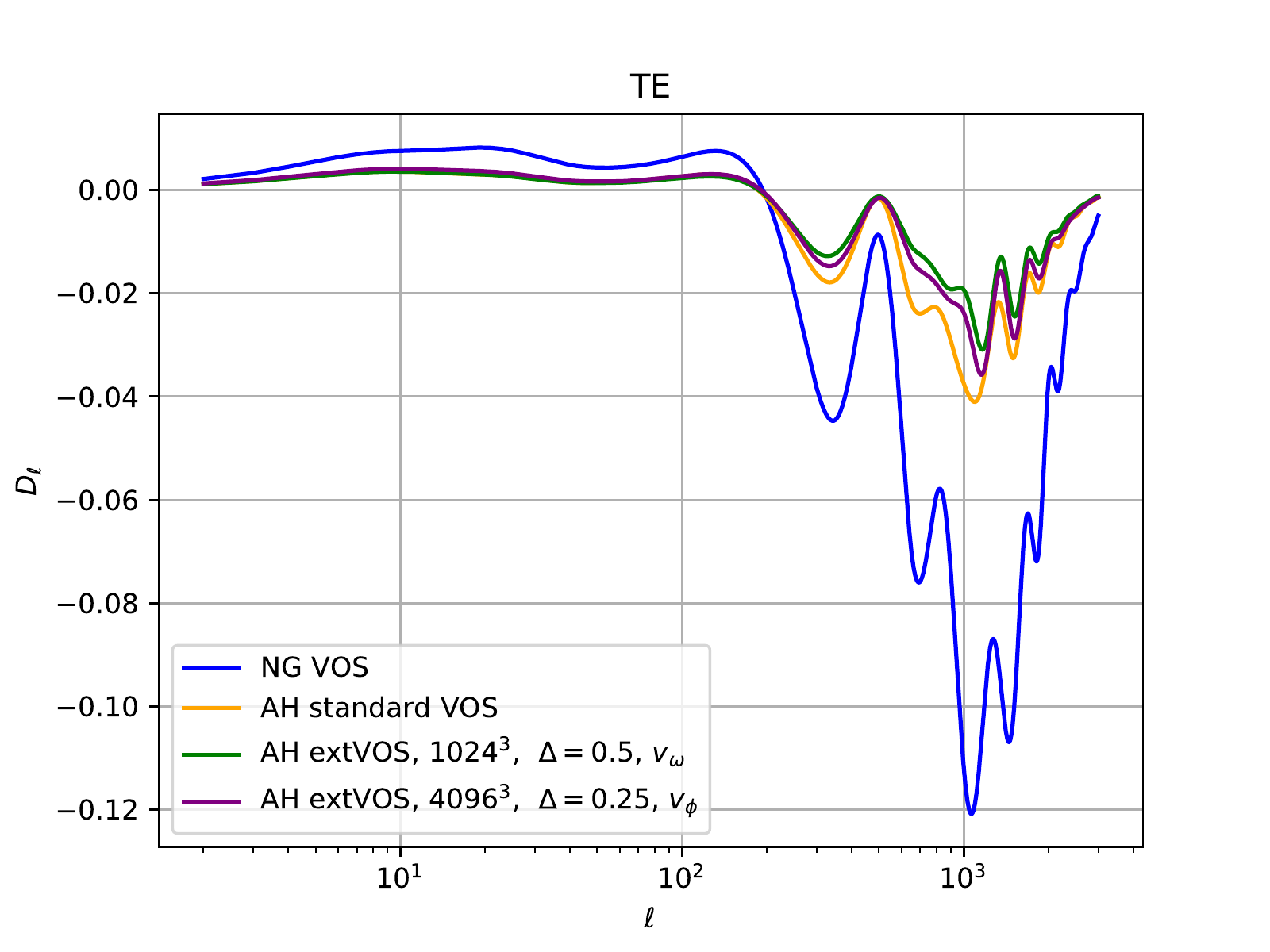}
  \includegraphics[width=1.0\columnwidth]{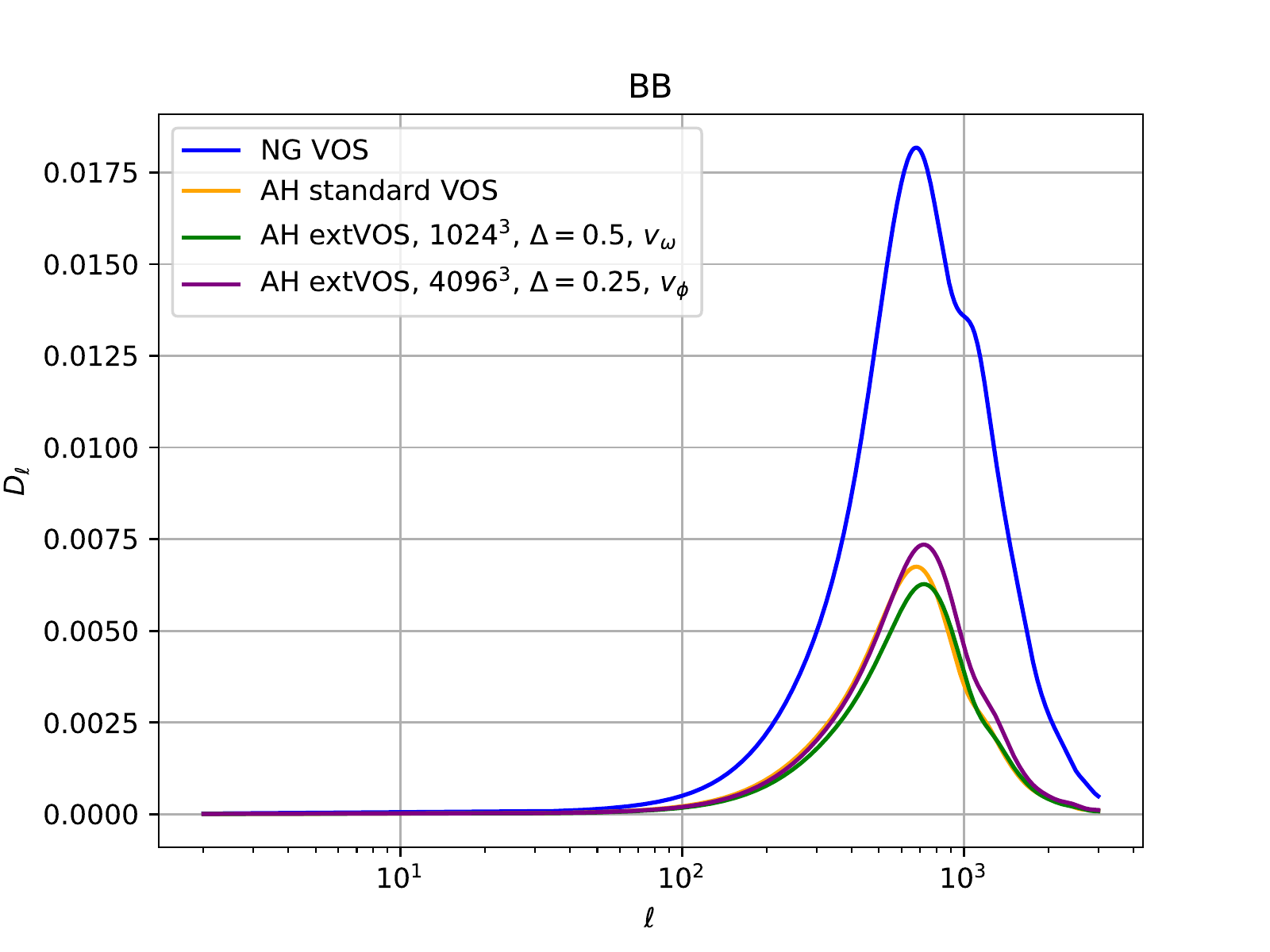}
  \caption{Power spectrum of cosmic microwave background anisotropies, obtained with the CMBACT4 code, for the standard Nambu-Goto calibration, the standard Abelian-Higgs calibration and two extended VOS calibrations in the present work. The panels depict the TT (top left), EE (top right), TE (bottom left) and BB (bottom right) spectra. In each case the spectrum is obtained by averaging over 200 realizations.}
  \label{fig9}
\end{figure*}

The obtained TT, EE, TE and BE spectra for the four calibrations can be seen in figure \ref{fig9}. We have conservatively normalized all four cases to a string tension $G\mu$ of $1.0 \times 10^{-7}$, even though the constraints on Nambu-Goto strings are stronger than those on Abelian-Higgs strings---the reason for this being clear from the figure itself. The fact that the obtained spectra for the three Abelian-Higgs cases are in better agreement with each other than with the Nambu-Goto case, is therefore to be expected. The key underlying reason for this difference can be ascribed to the different average values of the string velocities, which are larger in the Nambu-Goto case.

That being said, there are some significant differences between the three Abelian-Higgs calibrations. Comparing our most accurate calibration, at $4096^3$, with the standard VOS one, it should be noted that these differences are scale-dependent. They are larger at low-l for $TT$, and at high-l for $EE$, $BB$ and $BE$. Specifically, at the multipole $l=10$, the relative difference in the $TT$ power spectra is around $16\%$, $30\%$ and $11\%$ for the scalar, vector and tensor $C_l$, respectively. It is also worthy of note that among the three Abelian-Higgs calibrations the most discrepant is the $1024^3$ one relying on the equation of state velocity estimator. Admittedly, the extent to which these differences are entirely due to the different calibrations, rather than features due to the limitations of the unconnected segments model (which could plausibly introduce scale-dependent effects) is unclear. Nevertheless, this comparison highlights the need for accurate calibrations of the VOS model parameters.
 
\section{\label{conc}Conclusion}

In this work we have relied on our recently developed GPU-accelerated Abelian-Higgs cosmic string evolution code \cite{Correia:2018gew,Correia:2020yqg} to provide a more detailed and statistically robust calibration of the canonical velocity-dependent one-scale model, which updates and extends our previous analyses \cite{Correia:2019bdl,Correia:2020gkj}. Our data set of hundreds of high resolution simulations, comprising box sizes from $1024^3$ to $4006^3$ and exploring various choices of expansion rates, lattice spacing and velocity estimator, is by far the largest one carried out to date, and enables both a calibration of the model with small statistical uncertainties but also an assessment of the systematic uncertainties due to various numerical choices.

We have identified key differences between the equation of state and conjugate momentum estimators for the string velocities, depending on the resolution of the simulations, and have shown that the former one leads to unphysical results for fast expansion rates while the latter one is more reliable. It is interesting that this result is the opposite to the behaviour previously inferred in Minkowski space \cite{Hindmarsh:2017qff}. We emphasize that the two results are not mutually incompatible, since they pertain to opposite physical limits (fast expansion versus no expansion), and there is in any case no reason why a single numerical estimator algorithm should outperform all others in all physical settings.

Although the extended VOS model has 6 different model parameters (as opposed to only two parameters in the simpler version of the model) and there are some degeneracies among these parameters, which are clearly visible in our sets of corner plots, our analysis confirms that it is possible to accurately calibrate this extended model, with significant gains of physical insight into the dynamics of cosmic string networks. This is particularly the case when it comes to the energy loss mechanisms, and the relative roles of loop production and of scalar and gauge radiation. Unlike the case of domain walls where radiation losses always dominate \cite{Rybak1,Rybak2}, in this work we have confirmed that loop production and radiation losses both contribute, and indeed the production of large loops is clearly seen when one visualizes the evolution of large simulation boxes \cite{Movie1,Movie2}. Our work has shown that high resolution simulations, based on reliable estimators, are necessary for an accurate calibration. These effects in particular, and the detailed modelling provided by the extended VOS model in general, are important for credible forecasts of the constraints on cosmic strings to be expected from future experiments such as LISA and CORE \cite{LISA,CORE}.

Finally, we note that although in this work we restricted ourselves to box sizes up to $4096^3$, it is very possible to go further. Our highly efficient GPU accelerated string network evolution code, which has been shown to have almost perfect weak scaling \cite{Correia:2020yqg} enables $8192^3$ production runs on 4096 PizDaint GPUs to be executed in 33.2 minutes of wall clock time. Moreover,
in situ visualization advances have enabled us to reduce data outputs by about four orders of magnitude. Several opportunities therefore stem from here. On the one hand, an analogous programme of $8192^3$ simulations is viable, including a detailed set of small-scale network diagnostics which, in particular, enable a robust comparison between Nambu-Goto and Abelian-Higgs (field theory) codes. On the other hand, our code can be extended to characterize the evolution of more realistic defects, specifically those with additional degrees of freedom on the string worldsheet, such as wiggly strings and superconducting strings. Both of these directions are currently being explored.

\begin{acknowledgments}
This work was financed by FEDER---Fundo Europeu de Desenvolvimento Regional funds through the COMPETE 2020---Operational Programme for Competitiveness and Internationalisation (POCI), and by Portuguese funds through FCT - Funda\c c\~ao para a Ci\^encia e a Tecnologia in the framework of the project POCI-01-0145-FEDER-028987 and PTDC/FIS-AST/28987/2017. J.R.C. is supported by an FCT fellowship (SFRH/BD/130445/2017). We gratefully acknowledge the support of NVIDIA Corporation with the donation of the Quadro P5000 GPU used for this research.

We acknowledge PRACE for awarding us access to Piz Daint at CSCS, Switzerland, through Preparatory Access proposal 2010PA4610, Project Access proposal 2019204986 and Project Access proposal 2020225448. Technical support from Jean Favre at CSCS is gratefully acknowledged.
\end{acknowledgments}

\bibliography{artigo}
\end{document}